\documentclass[acmsmall]{acmart}
\AtBeginDocument{%
  }

\setcopyright{acmlicensed}
\copyrightyear{2018}
\acmYear{2018}
\acmDOI{XXXXXXX.XXXXXXX}
\acmConference[In Proceedings of the 35rd ACM SIGSOFT International Symposium on SoftwareTesting and Analysis (ISSTA ’26),]{Make sure to enter the correct
  conference title from your rights confirmation email}{June 03--05,
  2018}{Oakland, CA}
\acmISBN{978-1-4503-XXXX-X/2018/06}

\usepackage{tabularx}
\usepackage{amsmath}

\usepackage{amssymb}
\usepackage{amsfonts}
\usepackage{subcaption}
\usepackage{graphicx}
\usepackage{wasysym}
\usepackage{color,soul}
\usepackage{multirow}
\usepackage{changepage}
\usepackage{threeparttable}
\usepackage{booktabs}
\usepackage[dvipsnames]{xcolor}
\usepackage[many]{tcolorbox} % for summary box
\usepackage{fontawesome5} % for icons
\usepackage{colortbl}
\usepackage{mathtools}
\newtcolorbox{boxK}{
    top=2pt,
    bottom=2pt,
    left=2pt,
    right=2pt,
    boxrule = 0pt,
    toprule = 0pt, % top rule weight
}

\newcommand{\inc}[1]{\textcolor{red}{\fontsize{6pt}{6pt}\selectfont\textsuperscript{\textbf{+#1\%}}}}
\newcommand{\dec}[1]{\textcolor{ForestGreen}{\fontsize{6pt}{6pt}\selectfont\textsuperscript{\textbf{-#1\%}}}}
\soulregister{\cite}7 % 注册\cite命令
\soulregister{\citep}7 % 注册\citep命令
\soulregister{\citet}7 % 注册\citet命令
\soulregister{\ref}7 % 注册\ref命令
\soulregister{\pageref}7 % 注册\pageref命令
\definecolor{headergray}{RGB}{240, 240, 240}
\definecolor{sagegreen}{RGB}{242, 250, 242}

\begin{document}

\title{SAGE: Signal-Amplified Guided Embeddings for LLM-based Vulnerability Detection}

\author{Zhengyang Shan}
\orcid{0009-0001-9291-6416}
\email{202520912@mail.sdu.edu.cn}
\affiliation{%
  \institution{Shandong University}
  \city{Qingdao}
  \state{Shandong}
  \country{China}
}

\author{Xu Qian}
\orcid{0009-0007-4687-8790}
\affiliation{%
  \institution{Shandong University}
  \city{Qingdao}
  \state{Shandong}
  \country{China}}
\email{202515294@mail.sdu.edu.cn}

\author{Jiayun Xin}
\orcid{0009-0009-2992-2432}
\affiliation{%
  \institution{Shandong University}
  \city{Qingdao}
  \state{Shandong}
  \country{China}}
\email{202535317@mail.sdu.edu.cn}

\author{Minghui Xu}
\orcid{0000-0003-3675-3461}
\affiliation{%
  \institution{Quancheng Laboratory}
  \city{Jinan}
  \state{Shandong}
  \country{China}
}
\affiliation{%
 \institution{Shandong University}
 \city{Qingdao}
 \state{Shandong}
 \country{China}}
\email{mhxu@sdu.edu.cn}
\authornote{Corresponding author.} 

\author{Yue Zhang}
\orcid{0000-0002-7786-0231}
\affiliation{%
  \institution{Quancheng Laboratory}
  \city{Jinan}
  \state{Shandong}
  \country{China}
}
\affiliation{%
 \institution{Shandong University}
 \city{Qingdao}
 \state{Shandong}
 \country{China}}
\email{zyueinfosec@sdu.edu.cn}
\authornotemark[1]

\author{Zhen Yang}
\orcid{0000-0003-0670-4538}
\affiliation{%
  \institution{Shandong University}
  \city{Qingdao}
  \state{Shandong}
  \country{China}
}
\email{zhenyang@sdu.edu.cn}

\author{Hao Wu}
\orcid{0000-0002-0980-9805}
\email{hao.wu@nju.edu.cn}
\affiliation{%
  \institution{Nanjing University}
  \city{Nanjing}
  \state{Jiangsu}
  \country{China}
}

\author{Xiuzhen Cheng}
\orcid{0000-0001-9998-2398}
\affiliation{%
  \institution{Shandong University}
  \city{Qingdao}
  \state{Shandong}
  \country{China}
}
\email{xzcheng@sdu.edu.cn}

\renewcommand{\shortauthors}{Trovato et al.}

\begin{abstract}
Software vulnerabilities are a primary threat to modern infrastructure. While static analysis and Graph Neural Networks have long served as the foundation for vulnerability detection, the emergence of Large Language Models (LLMs) has introduced a transformative paradigm driven by superior semantic reasoning and cross-environment generalization. However, in the context of LLM-based vulnerability detection, we identify a fundamental bottleneck in these models termed \textbf{Signal Submersion}: a state where features related to vulnerability are activated internally but numerically overwhelmed by dominant functional semantics. To address this, we propose \textbf{SAGE} (\textbf{S}ignal-\textbf{A}mplified \textbf{G}uided \textbf{E}mbeddings), a framework that shifts from passive signal submersion to active signal recovery. SAGE integrates task-conditional Sparse Autoencoders (SAEs) to isolate and amplify these faint vulnerability signals. Extensive evaluations on BigVul, PrimeVul, and PreciseBugs demonstrate that SAGE achieves state-of-the-art performance. Notably, SAGE mitigates Signal Submersion by increasing the internal Signal-to-Noise Ratio (SNR) by 12.7$\times$ via sparse manifold projection. This mechanistic intervention enables a 7B model to achieve up to 318\% Matthews Correlation Coefficient (MCC) gains on unseen distributions and a 319\% gain on classic datasets. By maintaining robust performance across 13 programming languages and outperforming 34B baselines, SAGE establishes a more efficient and scalable path to software security than simple parameter scaling.
\end{abstract}



\maketitle

\section{Introduction}
The rapid expansion of the digital attack surface has led to a surge in software vulnerabilities, with global databases like the NIST National Vulnerability Database (NVD) and the MITRE CVE list consistently logging thousands of new entries monthly~\cite{NIST}. These flaws, often stemming from subtle coding oversights, can trigger a cascade of consequences--from localized data exfiltration and unauthorized privilege escalation to systemic compromises and catastrophic service disruptions~\cite{vuldect}. To combat this, detection methods have evolved through several distinct paradigms: starting with traditional static analysis~\cite{statis}, moving toward Graph Neural Networks (GNNs)~\cite{devign,vuldeepecker,sysevr,deeplearning}, and more recently, adopting pre-trained models such as CodeBERT and UniXcoder~\cite{feng-etal-2020-codebert,guo2022unixcoderunifiedcrossmodalpretraining}. 

The current state-of-the-art has shifted toward Large Language Models (LLMs), which represent a leap in vulnerability detection. Unlike previous iterations, LLMs possess sophisticated semantic comprehension and reasoning capabilities. This allows them to interpret the latent intent of code and trace complex execution logic across extensive sequences. By internalizing cross-lingual patterns and nuanced software environments without the need for manual rule crafting, LLMs offer a more generalized and robust solution to identifying vulnerabilities.

However, despite these impressive capabilities, even the most advanced models available today still face critical limits in their performance. Prior evaluations show that, on realistic vulnerability-detection benchmarks, frontier LLMs often perform substantially worse than their general coding ability would suggest, and may approach random guessing when the distinction between safe and vulnerable code is subtle and the label distribution is highly imbalanced~\cite{primevul,wen2025boostingvulnerabilitydetectionllms}. In many cases, these frontier models struggle to identify the tiny, crucial differences between safe and unsafe code, leading to results that are sometimes no better than random guessing~\cite{comprehensive,overfitting}. The primary reason for this failure is that the signals of a vulnerability are often very weak and its code always hidden inside hundreds of lines of normal, working code~\cite{subtle,dataquality}. Because the surrounding code looks perfectly correct and follows all the rules of the programming language, it creates a lot of background noise. Even with their advanced reasoning, LLMs find it hard to focus on the tiny parts of the code that cause a security risk~\cite{comprehensive,overfitting}.

We identify this phenomenon as \textbf{Signal Submersion}. Our analysis reveals that while features related to vulnerabilities are activated in LLMs, their activation intensity remains weak compared to the dominant global functional semantics. Consequently, these subtle signals become numerically dominated by the model's standard processing, creating a performance bottleneck where vulnerability signals in final layers are too weak to enable discriminative predictions.

Motivated by this, we propose \textbf{SAGE} (\textbf{S}ignal-\textbf{A}mplified \textbf{G}uided \textbf{E}mbeddings), a framework that transitions from passive signal submersion to active signal recovery. SAGE integrates task-conditional Sparse Autoencoders (SAEs) trained independently at each layer, optimizing the SAE and classifier via a joint loss function. This design explicitly constructs a sparse manifold to amplify vulnerability signals while suppressing the semantic background, capturing features at their peak magnitude before Signal Submersion occurs. Crucially, across the tested decoder-only backbones, SAGE can be integrated without modifying the backbone parameters. By skipping the need to run the full LLM during training, it achieves state-of-the-art results with faster speed.

We evaluated SAGE across four model architectures on refined versions of BigVul~\cite{bigvul}, PrimeVul~\cite{primevul}, and PreciseBugs~\cite{precisebugs}, benchmarking against 10 state-of-the-art methods and commercial frontier models including GPT-5.2-Codex~\cite{gpt52}, Gemini 3 pro~\cite{gemini3} and Claude Opus 4.5~\cite{claude45}. Results demonstrate that SAGE consistently outperforms baselines and exhibits superior robustness. Mechanistic analysis confirms our approach effectively amplifies the internal signal-to-noise ratio. Our contributions are summarized as follows:
\begin{itemize}
    \item \textbf{Identification of Signal Submersion:} We identify Signal Submersion as an important bottleneck for LLM-based vulnerability detection in the tested model families. Our empirical analysis reveals that vulnerability features are not absent but are numerically overwhelmed, constituting less than 5\% of the total activation magnitude in deep layers.
    \item \textbf{The SAGE Framework:} We propose SAGE, a framework that transitions from passive submersion to active amplification. By projecting entangled states onto a sparse manifold, SAGE reverses Signal Submersion, boosting the signal-to-noise ratio (SNR) by up to 12.7 times compared to standard inference.
    \item \textbf{State-of-the-Art Performance:} We conduct an extensive evaluation across refined security benchmarks, including PrimeVul, BigVul, and PreciseBugs. SAGE establishes state-of-the-art (SOTA) performance, delivering a \textbf{318\%} improvement in Matthews Correlation Coefficient (MCC) on unseen distributions and a \textbf{319\%} gain on classic datasets. Crucially, we demonstrate that a 7B model integrated with SAGE significantly outperforms 34B parameter baselines and even larger commercial frontier model, showing positive transfer across 13 programming languages, while results on low-resource language subsets should be interpreted cautiously. 
\end{itemize}

\section{Related Work}

\subsection{Traditional Vulnerability Detection Methods}
The field of vulnerability detection has undergone significant evolution, beginning with foundational static analysis tools ~\cite{Coverity, ezfindingbug, CPG} and rule-based systems ~\cite{pattern, javastaticanalysis}. These early approaches established the essential baseline for automated security auditing by encoding known vulnerability patterns. To move beyond syntax-level checks, dynamic techniques such as fuzzing and symbolic execution ~\cite{fuzzing} were introduced, significantly enhancing program state exploration and interaction depth. While effective in validating execution-time behaviors, these methods often encounter scalability challenges when navigating expansive path spaces or identifying flaws hidden under rare edge conditions ~\cite{empirical, mitigating, manifesto}.
To bridge the gap between efficiency and deep semantic understanding, subsequent research has shifted towards learning-based paradigms. Recent advancements leverage graph neural networks to operate on structured representations, such as Code Property Graphs ~\cite{deeplearning} or Program Dependence Graphs ~\cite{Li_2021}. By capturing long-range dependencies and intricate control-flow relationships, these structure-based methods have demonstrated a superior ability to model the complex semantics inherent in modern software vulnerabilities.

\subsection{LLM-based Vulnerability Detection}
\label{sec:llm-base}
\textbf{Prompt-based Methods.} 
Zero-shot and few-shot prompting strategies leverage the generalization capabilities of LLMs without the need for extensive weight updates ~\cite{wu2023exploringlimitschatgptsoftware}. Research has explored various prompt engineering techniques ~\cite{zhang2024promptenhancedsoftwarevulnerabilitydetection} and established unified evaluation frameworks ~\cite{sun2025llm4vulnunifiedevaluationframework} to assess these capabilities. Studies indicate that while LLMs show promise, they primarily rely on token-level patterns, and distinguishing subtle vulnerabilities remains a complex challenge ~\cite{zhang2023blackboxattackcodemodels}. Comprehensive evaluations suggest that standard prompting strategies face difficulties in achieving high precision on complex real-world datasets, indicating the complexity of the task ~\cite{comprehensive}.

\textbf{Fine-tuning-based Methods.} 
Fine-tuning pre-trained models, such as LineVul ~\cite{linevul} and CodeBERT ~\cite{feng-etal-2020-codebert}, marked a step forward in detection performance. Following these advancements, the community has increasingly focused on the robustness of evaluation protocols. Recent studies note that random splitting on historical datasets~\cite{bigvul} can lead to data leakage, where models might retain specific sample features rather than learning generalizable patterns ~\cite{overfitting, dataquality, revisiting}. Similarly, investigations into models like PDBERT ~\cite{pdbert} emphasize the importance of distinct training and testing distributions to ensure genuine comprehension. Addressing these insights, recent works like VulLLM have adopted multi-task instruction tuning ~\cite{du-etal-2024-generalization} and strict temporal splitting protocols ~\cite{precisebugs, yang2024securityvulnerabilitydetectionmultitask} (e.g., MSIVD) to better align model training with realistic, evolving vulnerability landscapes.

\textbf{Reinforcement Learning-based Methods.} 
To align models with non-differentiable security objectives, researchers have integrated Reinforcement Learning (RL). Methodologies range from optimizing prompts via policy gradients ~\cite{ren2024prorlearn} to continuous prefix tuning for generating secure code ~\cite{he2023large, shojaee2023execution}. More recently, efforts have shifted towards internalizing reasoning capabilities. Inspired by advancements in reasoning models ~\cite{guo2025deepseek}, approaches like those by Simoni et al. ~\cite{simoni2025improving} utilize Group Relative Policy Optimization (GRPO) to enhance detection logic. Frameworks such as ReVD ~\cite{wen2025boostingvulnerabilitydetectionllms} further employ curriculum learning and synthetic reasoning traces to improve the model's sensitivity to subtle code distinctions.

\textbf{Agent-based Methods.} 
Multi-Agent Systems (MAS) have been proposed to simulate collaborative auditing processes. Frameworks like MuCoLD ~\cite{mao2024multiroleconsensusllmsdiscussions} utilize multi-role debates to refine consensus, while PentestGPT ~\cite{deng2024pentestgptllmempoweredautomaticpenetration} facilitates automated penetration testing. To address the computational overhead of multi-agent interactions, adaptive frameworks like ConColl ~\cite{concoll} dynamically balance single-agent efficiency with multi-agent collaboration. Furthermore, systems such as AutoCodeRover ~\cite{zhang2024autocoderoverautonomousprogramimprovement} extend these capabilities beyond detection, exploring autonomous program repair.

\subsection{Internal Representations for Vulnerability Analysis}
Complementing external prompting and fine-tuning, mechanistic interpretability analyzes internal LLM states to understand code processing. Research indicates that attention mechanisms inherently encode both syntactic structures (ASTs) ~\cite{Wan2022} and semantic dependencies (CFGs) ~\cite{Ma2024}, enabling vulnerability localization methods that leverage these internal signals. Building on the linear representation hypothesis ~\cite{Park2023}, activation-steering methods modify model behavior by adding dense steering vectors to hidden states during inference ~\cite{Li2025Steering}. These approaches are lightweight and effective when a single global direction is sufficient to bias model behavior. However, for vulnerability detection, the target signal is often sparse, weak, and entangled with dominant functional semantics. In this setting, a fixed dense intervention may not adequately separate vulnerability-specific cues from the broader semantic background. SAGE differs from steering-based methods in two ways: it learns a sparse, task-conditional representation rather than applying a fixed dense offset, and it explicitly selects intermediate layers where the vulnerability signal is most separable before deep-layer dilution occurs. Recently, Sparse Autoencoders (SAEs) ~\cite{Bricken2023} have emerged as a powerful tool for disentangling polysemantic features. While Melo et al. ~\cite{Melo2025} pioneered the application of SAEs to bug detection using general-purpose models ~\cite{Lieberum2024}, there remains an opportunity to further align these sparse representations with the specific, fine-grained semantics required for security tasks. This motivates the exploration of task-specific optimization for internal feature extraction.

In summary, while existing approaches have evolved from static analysis to advanced deep learning, they share limitations in handling the subtle nature of vulnerability signals. Table \ref{tab:related_work} provides a systematic comparison of these paradigms. Current vulnerability detection methods mainly rely on parsing code graphs or optimizing model weights through fine-tuning and reinforcement learning. However, these approaches either rely on heavy external priors or succumb to signal submersion by optimizing on diluted final-layer representations.
SAGE works as an orthogonal framework that complements these existing methods. Instead of updating the model parameters, SAGE keeps the backbone frozen and intervenes directly at the intermediate layers. This approach is not limited by the model's previous training or alignment history. By focusing on these internal layers, we can find and use hidden features that are usually lost during standard end-to-end training.

\begin{table}[h] % 建议使用 [t] 让表格浮动到页顶，通常比 [h] 更节省正文流空间
\centering
\caption{Comparison of Technical Paradigms. \textbf{SAGE} is an orthogonal framework enhancing internal representations, compatible with SFT, RL, and Agents.}
\label{tab:related_work}
\vspace{-0.8em} % 调整 caption 和表格的间距

\footnotesize 
\renewcommand{\arraystretch}{1.1} % 将行高从 1.2 压缩到 1.05
\setlength{\tabcolsep}{3pt}      % 减小列间距，让文字有更多横向空间

% 改为两列布局：Category (含Works) | Core Mechanism
\begin{tabularx}{\linewidth}{l X}
\toprule
\rowcolor{headergray} 
\textbf{Category \& Rep. Works} & \textbf{Core Mechanism} \\ 
\midrule

% Graph Based
\begin{tabular}[t]{@{}l@{}} \textbf{Graph-based} \\ \scriptsize ~\cite{deeplearning, Li_2021} \end{tabular} & 
\textbf{Explicit Structure Modeling.} Utilizes Graph Neural Networks (GNNs) to learn from intermediate representations like Code Property Graphs (CPG) or PDGs. \\ 
\midrule

% SFT
\begin{tabular}[t]{@{}l@{}} \textbf{Supervised Fine-Tuning} \\ \scriptsize ~\cite{linevul, yang2024securityvulnerabilitydetectionmultitask} \end{tabular} & 
\textbf{Weight Optimization.} Updates global model parameters via end-to-end backpropagation to align final-layer probability distributions with ground truth. \\ 
\midrule

% RL
\begin{tabular}[t]{@{}l@{}} \textbf{Reinforcement Learning} \\ \scriptsize ~\cite{wen2025boostingvulnerabilitydetectionllms} \end{tabular} & 
\textbf{Policy Alignment.} Optimizes the reasoning generation policy using reward signals (e.g., COPO/GRPO) derived from correct detection outcomes. \\ 
\midrule

% Multi-Agent
\begin{tabular}[t]{@{}l@{}} \textbf{Agent-based} \\ \scriptsize ~\cite{mao2024multiroleconsensusllmsdiscussions, concoll} \end{tabular} & 
\textbf{Workflow Orchestration.} Simulates multi-role debates or sequential retrieval-augmented workflows to decompose complex detection tasks. \\ 
\midrule

% Steering
\begin{tabular}[t]{@{}l@{}} \textbf{Activation Engineering} \\ \scriptsize ~\cite{Li2025Steering} \end{tabular} & 
\textbf{Dense Intervention.} Modifies behavior by adding a fixed, dense steering vector to the activation space during inference. \\ 
\midrule

% Dictionary Learning
\begin{tabular}[t]{@{}l@{}} \textbf{Dict. Learning} \\ \scriptsize ~\cite{Melo2025} \end{tabular} & 
\textbf{Unsupervised Extraction.} Decomposes activations into sparse features using Autoencoders (e.g., Gemma Scope) on general corpora. \\ 
\midrule

% SAGE (Ours)
\rowcolor{sagegreen} 
\textbf{SAGE (Ours)} & 
\textbf{Pan-Layer Perception + Task-Conditional SAE.} Relation: Orthogonal to SFT/RL (operates with frozen backbone parameters, compatible with any model regardless of its alignment history) and Complementary to Agents (enhancing atomic capability). \\ 
\bottomrule
\end{tabularx}
\end{table}

\section{Method}
As we illustrated, the core challenge in vulnerability detection is Signal Submersion, where critical intermediate features are progressively overwhelmed by global semantics. Critically, post-training approaches ~\cite{linevul,yang2024securityvulnerabilitydetectionmultitask,wen2025boostingvulnerabilitydetectionllms,simoni2025improving} do not explicitly model the internal processing. By ignoring the evolution of intermediate features and calculating loss on representations where signals are already diluted, these methods may rely on spurious correlations, leading to poor generalization. To overcome this, SAGE actively intervenes in the model's intermediate information flow. By projecting entangled activations into a sparse manifold via a task-conditional Sparse Autoencoder (SAE), SAGE explicitly disentangles and amplifies faint vulnerability signals. This section begins by formally defining the Signal Submersion problem, followed by a detailed technical breakdown of SAGE’s three core components: Pan-Layer Latent Perception, Sparse Manifold Projection, and Task-Conditional Latent Alignment.

\subsection{Motivation and Problem Formulation}
\label{sec:problem_formulation}

Formally, let $\mathcal{D} = \{(x_i, y_i)\}_{i=1}^N$ denote the dataset, where $x$ is the source code and $y \in \{0, 1\}$ is the label ($0$ for safe, $1$ for vulnerable). For a Transformer-based LLM with $L$ layers, let $h_l(x) \in \mathbb{R}^d$ represent the latent activation of input $x$ at layer $l$.

We first categorize the activations at layer $l$ into two sets based on their labels: $H_{safe}^{(l)} = \{h_l(x) \mid (x, y) \in \mathcal{D}, y=0\}$ and $H_{vuln}^{(l)} = \{h_l(x) \mid (x, y) \in \mathcal{D}, y=1\}$. Let $H_{all}^{(l)} = H_{safe}^{(l)} \cup H_{vuln}^{(l)}$ denote the entire population of activations.
Drawing on the Linear Representation Hypothesis~\cite{Park2023,Li2025Steering}, we introduce two \emph{dataset-level geometric statistics} at layer $l$: a semantic centroid $s_l$ and a vulnerability direction $v_l$. Importantly, we do \emph{not} regard these quantities as an exact sample-wise causal decomposition of every activation. Instead, they define a descriptive coordinate system for characterizing how vulnerability-relevant information is organized relative to the dominant semantic background:
\[
s_l \triangleq \mathrm{Mean}\!\left(H_{\text{all}}^{(l)}\right),
\qquad
v_l \triangleq \mathrm{Mean}\!\left(H_{\text{vuln}}^{(l)}\right) - \mathrm{Mean}\!\left(H_{\text{safe}}^{(l)}\right).
\]

For an individual sample $x$, we express its activation relative to this coordinate system as
\begin{equation}
h_l(x) = s_l + \alpha_l(x)\, v_l + r_l(x),
\label{eq:geom_decomp}
\end{equation}
where
\[
\alpha_l(x)=\frac{\langle h_l(x)-s_l,\; v_l\rangle}{\|v_l\|_2^2}
\]
measures the alignment of sample $x$ with the vulnerability direction, and $r_l(x)$ denotes the residual component not explained by this direction. Under this formulation, $s_l$ represents the dominant semantic background shared across samples, while $\alpha_l(x)v_l$ captures the class-discriminative component aligned with vulnerability-related variation.

This formulation is descriptive rather than causal: throughout the paper, we use the term ``vulnerability signal'' to refer to the discriminative component aligned with $v_l$, not to a uniquely identifiable latent factor. 

\begin{figure}[h]
    \includegraphics[width=1\linewidth]{figure1b.pdf}
    \vspace{-1em}
    \caption{\textbf{Illustration of Signal Submersion and SAGE.} 
The self-attention mechanism assigns dominant weights to global functional context ($s_l$), causing the sparse vulnerability features ($v_l$) to be diluted during aggregation. 
SAGE addresses this by using a Sparse Autoencoder (SAE) to amplify the submerged vulnerability signal from the functional background.}
    \label{fig:main}
    \vspace{-0.7em}
\end{figure}

\textbf{The Signal Submersion Phenomenon.} 
While $v_l$ may be distinct in shallow layers, it suffers from a dilution effect as depth $l$ increases. As illustrated in Fig.~\ref{fig:main}, the self-attention mechanism aggregates long-range functional context. Even if a specific token captures a defect, its contribution is numerically suppressed by the massive weight of global functional tokens during the weighted sum calculation. 
Consequently, the vulnerability signal becomes negligible compared to the accumulating functional background in the final layer:
\begin{equation}
||s_L||_2 \gg ||v_L||_2.
\label{eq:submersion}
\end{equation}
Our empirical analysis confirms this, showing the relative magnitude of $v_l$ drops below 5\% in deeper layers.

This inequality reveals a bottleneck for conventional methods that rely on the final representation $h_L$. 
Since the discriminative component aligned with $\mathbf{v}_L$ is submerged by the dominant semantic component centered around $\mathbf{s}_L$, the optimization objective degenerates. Models effectively either performance degrade to a majority-class prior (ignoring the weak $v_L$) or overfit to stylistic heuristics within $s_L$ as proxies for vulnerability.
This necessitates our SAGE framework, which actively intervenes at an intermediate layer $l^*$ to recover $v_l$ before the Signal-to-Noise Ratio (SNR) deteriorates.

\subsection{Pan-Layer Latent Perception}\label{sec:pan_layer}

\textbf{Process-Based Perception via Intermediate Intervention.}
The Signal Submersion analysis reveals that the final representation $h_L$ operates in a structurally low SNR regime. We actively intervene in the model's inference trajectory, extracting latent states at intermediate layers ($l < L$) to recover vulnerability artifacts before they are submerged by global semantics.

\textbf{Hidden State Extraction.}
To intervene effectively, we must aggregate the sequence of hidden states $H^{(l)}$ into a representative vector $z^{(l)}$. We leverage the autoregressive properties of decoder-only large language models to extract the activation state of the last token in a token sequence. The causal attention mask ensures that the last token is the unique position that mathematically aggregates the receptive field of the entire code context $\{t_1, \dots, t_T\}$. Accordingly, we define the layer-wise representation as:
\begin{equation}
z^{(l)} = \mathcal{T}(H^{(l)}) \triangleq h_{t}^{(l)} \Big|_{t=T}.
\end{equation}
This operator preserves the cumulative reasoning trajectory encoded by the causal chain. While other aggregation methods (e.g., Mean Pooling, Max Pooling) exist, we prioritize the last token to avoid smoothing out high-frequency anomaly signals. A detailed empirical comparison demonstrating the superiority of this extraction strategy over others is provided in Section \ref{sec:rq2}.

\textbf{Layer-wise Independent Probing.}
Analyzing the model state at a specific layer is insufficient because the signals indicating a vulnerability vary across different depths. These signals may be strong in some layers but weak in others. To address this, SAGE employs task-conditional Sparse Autoencoders (SAEs) to analyze the representation $z^{(l)}$ at each layer individually. Because the backbone is frozen, these layer-wise SAEs are trained on cached hidden states rather than through repeated end-to-end backpropagation. The overhead therefore scales linearly with the number of probed layers, but avoids the substantially higher cost of updating the full backbone. This approach allows SAGE to scan through the model and identify exactly which layer provides the clearest and most separable vulnerability features.

\subsection{Sparse Manifold Projection via SAE}
\label{sec:sparse_projection}

\textbf{Separating Vulnerabilities.}
While the extracted latent states $z^{(l)}$ record the inference process, they represent multiple different concepts mixed together within a dense vector~\cite{Bricken2023}.
In this dense representation, strong features such as syntax often dominate the weaker signals associated with vulnerabilities.
Consequently, the vulnerability features are not readily separable using linear methods.
To address this, we hypothesize that mapping the state into a space with significantly more dimensions ($d_{sae} \gg d_{in}$) will separate these mixed features.
We employ an SAE to isolate the vulnerability signal as an independent and distinct feature.

\textbf{JumpReLU for Magnitude Preservation.}
Standard sparse coding methods typically rely on $L_1$ regularization.
However, this approach tends to reduce the value of all activations to achieve sparsity.
Since the vulnerability signals are already weak (Eq.~\ref{eq:submersion}), this reduction may reduce important information along with the noise.
To address this limitation, we utilize a JumpReLU SAE~\cite{deepmind2024jumping}.
Let $h_l \in \mathbb{R}^d$ be the input activation vector.
We project $h_l$ using a learnable encoder weight matrix $W_{enc} \in \mathbb{R}^{d_{sae} \times d}$ and a bias vector $b_{enc} \in \mathbb{R}^{d_{sae}}$, followed by a thresholding operation:
\begin{equation}
z = \sigma_{\theta}(W_{enc}h_l + b_{enc}), \quad \text{where } \sigma_{\theta}(u)_i = u_i \cdot \mathbb{I}(u_i > \theta_i).
\label{eq:jumprelu_mechanism}
\end{equation}
Here, $\theta$ is a learnable threshold vector that acts as a filter.
This mechanism ensures magnitude preservation.
Once a feature $z_i$ exceeds its threshold $\theta_i$, its original value is kept unchanged.
Consequently, SAGE effectively filters out noise by setting it to zero without weakening the retrieved vulnerability signal.

\textbf{Optimization Objective.}
We train the SAE to minimize a combined loss function.
This function balances two goals: accurately reconstructing the input activation $h_l$ and maintaining sparsity.
Unlike standard methods that penalize the magnitude of activations, we aim to minimize the number of active features.
Since directly counting non-zero elements is difficult for optimization, we use a sigmoid function $\sigma_{sig}$ to approximate this count:
\begin{equation}
\mathcal{L}_{SAE} = \underbrace{||h_l - W_{dec} z||_2^2}_{\mathcal{L}_{recon}} + \lambda \cdot \underbrace{\frac{1}{d_{sae}} \sum_{i=1}^{d_{sae}} \sigma_{sig}\left(\frac{u_i - \theta_i}{\tau}\right)}_{\mathcal{L}_{sparse} (\approx L_0)}.
\label{eq:loss_total}
\end{equation}
Here, $W_{dec} \in \mathbb{R}^{d \times d_{sae}}$ is the decoder weight matrix that maps the sparse feature $z$ back to the original dimension.
The parameter $\tau$ controls how sharply the function distinguishes between active and inactive features.
By minimizing $\mathcal{L}_{sparse}$, SAGE suppresses noise and improves the SNR. 

However, it is important to note that this optimization is unsupervised. The model aims to reconstruct the dominant semantic structure of the code and does not prioritize vulnerability-specific features. Since vulnerability signals are often faint and appear infrequently, the model tends to discard them to minimize the overall reconstruction error. Consequently, the standard SAE may fail to isolate these critical signals. To address this limitation, we introduce Task-Conditional Latent Alignment.
\subsection{Task-Conditional Latent Alignment}
\label{sec:task_alignment}

\textbf{Gradient Injection via Weighted Supervision.}
To transform the SAE from a passive feature extractor into a guided discovery engine, we introduce \textbf{Task-Conditional Latent Alignment}. We attach a linear probe $g_{\phi}$ directly to the sparse feature $z$. Crucially, to address the extreme class imbalance where safe samples vastly outnumber vulnerabilities, we employ Inverse-Frequency Risk Minimization. We formulate the supervision as a Weighted Cross-Entropy loss:
\begin{equation}
\mathcal{L}_{class} = - \frac{1}{B} \sum_{i=1}^{B} \sum_{c=1}^{C} \omega_c \cdot y_{i,c} \log(\hat{y}_{i,c}),
\end{equation}
where $n_c$ denotes the number of training samples in class $c$, and
\[
\omega_c = \frac{C \cdot (1/n_c)}{\sum_{j=1}^{C} (1/n_j)}
\]
is the normalized inverse-frequency weight for class $c$, with $C$ being the number of classes. This normalization ensures that the average class weight equals $1$, so that minority classes receive larger gradients without changing the overall loss scale.

We integrate this objective into a unified optimization framework:
\begin{equation}
\mathcal{L}_{total} = \underbrace{\mathcal{L}_{recon} + \lambda \cdot \mathcal{L}_{sparse}}_{\text{Unsupervised Structure}} + \gamma \cdot \underbrace{\mathcal{L}_{class}}_{\mathclap{\text{Supervised Guidance}}}.
\end{equation}
This architecture establishes a direct connection for gradient flow.
During backpropagation, the weighted gradients $\gamma \nabla \mathcal{L}_{class}$ from the classifier are propagated back to the SAE encoder. This process guides the JumpReLU activation mechanism. Unlike standard unsupervised learning, which separates all features indiscriminately, these gradients provide a supervised signal. These gradients guide the encoder to prioritize vulnerability patterns, regardless of their signal strength.
At the same time, the model is encouraged to down-weight strong but task-irrelevant features, such as coding style. We do not claim that this fully eliminates shortcut learning. Rather, SAGE reduces this risk through three design choices: strict de-duplication and temporal partitioning in the data pipeline, a sparse bottleneck that penalizes diffuse correlations, and evaluation on distribution-shifted datasets where superficial cues transfer poorly.~\cite{lataguard2025}.

\section{Experimental Design}
In this section, we present a comprehensive evaluation of the SAGE framework, utilizing refined and deduplicated versions of three rigorously curated benchmarks: BigVul ~\cite{bigvul}, PrimeVul ~\cite{primevul}, and PreciseBugs ~\cite{precisebugs}. Our experimental design extends beyond standard performance metrics; we aim to empirically validate the existence of the Signal Submersion phenomenon and demonstrate SAGE’s capability to recover these latent vulnerability signals.
Specifically, we structure our analysis around four Research Questions (RQs) that progress from macroscopic performance comparisons to microscopic mechanistic interpretations, concluding with an investigation into the relationship between general coding capabilities and security sensitivity:

\textbf{RQ1 (Overall Effectiveness): How does SAGE perform on standard vulnerability detection benchmarks compared to state-of-the-art baselines?}
Given the challenge of detecting minute semantic deviations within complex code, we establish a rigorous comparison against diverse baselines, ranging from graph-based methods to recent LLMs. This validates whether SAGE successfully amplifies critical vulnerability signals to achieve superior detection accuracy.

\textbf{RQ2 (Mechanistic Verification): Does SAGE quantitatively and qualitatively mitigate Signal Submersion via robust feature decoupling?}
To validate the structural grounding of our approach, we analyze latent space properties through Internal Signal-to-Noise Ratio (SNR) and Top-$K$ feature attribution. These analyses confirm that SAGE effectively recovers submerged signals via robust feature decoupling rather than relying on dense noise.

\textbf{RQ3 (Component Efficacy \& Stability): How do different components contribute to SAGE, and is the performance robust?}
We assess the necessity of key components through a comprehensive ablation study, isolating the contributions of Sparse Manifold Projection and task-alignment modules. Additionally, we conduct hyperparameter sensitivity analysis to ensure that the reported performance gains are robust and stable across varying configurations.

\textbf{RQ4 (Impact of Model Scale): How does the Signal Submersion effect and SAGE’s effectiveness evolve across different model scales?}
We investigate whether the Signal Submersion effect persists or diminishes as model parameters and complexity increase. This analysis determines if SAGE remains essential for large-scale foundation models or if scaling naturally resolves the dominance of global semantics.

\subsection{Studied LLMs}
Table \ref{tab:studied_llms} lists our studied LLMs with their categories, model architectures, release timelines, parameter sizes, and effective context window limits. To ensure a comprehensive evaluation across the current landscape of LLMs, we systematically selected representative models from three distinct categories: (1) General-Purpose Open-Weights LLMs (i.e., Mistral-7B~\cite{mistral7b} and Llama-3.1-8B~\cite{llama31}), which serve as strong baselines for generalized reasoning; (2) Code-Specialized LLMs (i.e., the CodeLlama family~\cite{codellama} and Qwen2.5-Coder family~\cite{qwen25}), which are explicitly optimized for programming tasks; and (3) Proprietary Frontier Models (i.e., GPT-5.2-Codex~\cite{gpt52}, Gemini 3 Pro~\cite{gemini3}, and Claude Opus 4.5~\cite{claude45}), which represent the current state-of-the-art in closed-source commercial performance.

\begin{table}[h]
\setlength\tabcolsep{3pt} % 稍微减小列间距以适应更多内容
\vspace{-0.5em}
\setlength{\abovecaptionskip}{0cm}
\caption{Studied LLMs in this paper}
\label{tab:studied_llms}
\scriptsize
\begin{threeparttable}
\begin{tabular*}{\linewidth}{l@{\extracolsep{\fill}}cccccc}
\toprule
Category                     & Model                          & Model Type       & Time    & Size & Training Base & In/Out (Tokens) \\ \midrule
\multirow{3}{*}{Commercial}  & GPT-5.2-Codex ~\cite{gpt52}           & Instruction Model & 2025-12 & /  & /             & 400k / 128k     \\ 
                             & Gemini 3 Pro ~\cite{gemini3}    & Instruction Model & 2025-11 & /    & /             & 1M / 64k       \\
                             & Claude Opus 4.5 ~\cite{claude45}& Instruction Model & 2025-11 & /    & /             & 200k / 64k     \\ \midrule
\multirow{2}{*}{General}     & Llama-3.1-8B ~\cite{llama31}    & Instruction Model & 2024-07 & 8B   & 15 trillion   & 128k / --     \\
                             & Mistral-7B ~\cite{mistral7b}    & Base Model       & 2023-10 & 7B   & /             & 32k / --        \\ \midrule
\multirow{4}{*}{Code}        & CodeLlama-7B ~\cite{codellama}  & Instruction Model & 2023-08 & 7B   & 500 billion   & 16k / --        \\
                             & CodeLlama-13B ~\cite{codellama} & Instruction Model & 2023-08 & 13B  & 500 billion   & 16k / --        \\ 
                             & CodeLlama-34B ~\cite{codellama} & Instruction Model & 2023-08 & 34B  & 500 billion   & 16k / --        \\
                             & Qwen2.5-Coder-7B ~\cite{qwen25}     & Instruction Model & 2024-09 & 7B   & 5.5 trillion  & 128k / --     \\ 
                             & Qwen2.5-Coder-14B ~\cite{qwen25}    & Instruction Model & 2024-09 & 14B  & 5.5 trillion  & 128k / --      \\ 
                             & Qwen2.5-Coder-32B ~\cite{qwen25}    & Instruction Model & 2024-11 & 32B  & 5.5 trillion  & 128k / --      \\ 
\bottomrule
\end{tabular*}
\begin{tablenotes}[flushleft]
\definecolor{light cyan}{HTML}{E1FFFF}
\fboxsep1.5pt
\item[$\Phi$] \scriptsize {In the Training Base column, ``/'' denotes undisclosed proprietary data. For CodeLlama, the 500 billion tokens represent code-specific training on top of the Llama 2 base. For open-weights models, the maximum output length is dynamically determined by the remaining context window. Therefore, we report the total context limit as the input capacity and mark the output as ``--'' to indicate this dynamic dependency.}
\end{tablenotes}
\end{threeparttable}
\end{table}
\vspace{-1em}

\subsection{Datasets and Preprocessing}
\label{sec:exp_design_data}

To ensure ecological validity and mitigate data leakage~\cite{dataquality,revisiting}, we employ distinct preprocessing protocols to curate refined versions of BigVul~\cite{bigvul}, PrimeVul~\cite{primevul}, and PreciseBugs~\cite{precisebugs}.

\textbf{Fuzzy De-duplication Protocol.} 
Since BigVul and PrimeVul are historical snapshots likely present in LLM pre-training corpora, standard splitting is ineffective. To counter memorization, we implement a strict semantic filtering mechanism. We define an abstraction mapping $\psi$ to neutralize trivial perturbations and utilize SimHash with Jaccard coefficient $\mathcal{J}$ for efficient retrieval. We construct a purified training set $D_{train}$ by enforcing a similarity threshold $\tau = 0.75$ against the test set:
\begin{equation}
D_{train} = \{ x \in D_{train}^{raw} \mid \forall y \in D_{test}, \mathcal{J}(\psi(x), \psi(y)) < \tau \},
\end{equation}
where $\mathcal{J}(A, B) = \frac{|A \cap B|}{|A \cup B|}$. This ensures no training sample is semantically isomorphic to any held-out test sample.

\textbf{Temporal Partitioning.} 
For the continuously updated PreciseBugs, we adopt a temporal splitting protocol~\cite{yang2024securityvulnerabilitydetectionmultitask} to simulate zero-day detection. We set a global cutoff $T_{cutoff}$ at January 2023, aligning with standard baselines like CodeLlama~\cite{codellama} and strictly predating evaluation samples for most studied models. The evaluation set comprises post-cutoff samples:
\begin{equation}
D_{test}^{PB} = \{ x \in D_{total}^{PB} \mid \text{Time}(x) > T_{cutoff} \}.
\end{equation}
This formulation minimizes memorization bias and establishes a consistent benchmark across model generations.

\textbf{Dataset Statistics and Diversity.}
Table~\ref{tab:dataset_stats} summarizes the statistics of the processed datasets. 
Regarding linguistic diversity, while BigVul and PrimeVul predominantly feature C/C++ vulnerabilities typical of systems software, PreciseBugs is inherently multi-lingual (comprising C/C++, Java, Python, and JavaScript). 
This inclusion allows us to evaluate SAGE's cross-language generalization capabilities beyond the single-language limitations of traditional benchmarks.

\begin{table*}[h]
\centering
\caption{Detailed statistics of the refined datasets. To conserve space, complexity metrics (Tokens, LoC, CCN) are reported as Mean $\pm$ Std, excluding median/p95 values.}
\vspace{-1em}
\label{tab:dataset_stats}
\resizebox{1.0\textwidth}{!}{%
\begin{tabular}{llccccc}
\toprule
\textbf{Dataset} & \textbf{Lang.} & \textbf{Samples} & \textbf{Balance Ratio} & \textbf{Tokens} & \textbf{LoC} & \textbf{CCN} \\
 & & (Train / Test) & (Pos:Neg Train/Test) & (Mean $\pm$ Std) & (Mean $\pm$ Std) & (Mean $\pm$ Std) \\
\midrule

% BigVul
\textbf{BigVul} & C/C++ & 16296 / 18864 & 1:1 / 1:17 & 471.3 $\pm$ 1242.7 & 40.7 $\pm$ 110.4 & 6.7 $\pm$ 12.6 \\

% PrimeVul
\textbf{PrimeVul} & C/C++ & 9157 / 24788 & 1:1 / 1:44 & 693.0 $\pm$ 2459.3 & 57.1 $\pm$ 206.4 & 8.9 $\pm$ 16.4 \\
\midrule

% PreciseBugs Breakdown
\textbf{PreciseBugs} & & & & & & \\
\hspace{1em} \textit{C\#} & C\# & 111 / 53 & 1:4 / 1:2 & 4830.7 $\pm$ 10247.9 & 405.5 $\pm$ 587.0 & 1.4 $\pm$ 1.4 \\
\hspace{1em} \textit{C/C++} & C/C++ & 5957 / 315 & 1:4 / 1:4 & 17883.9 $\pm$ 16986.8 & 1507.1 $\pm$ 1366.2 & 2.8 $\pm$ 7.9 \\
\hspace{1em} \textit{Go} & Go & 1110 / 181 & 1:4 / 1:4 & 5629.4 $\pm$ 9218.4 & 459.6 $\pm$ 694.4 & 3.1 $\pm$ 4.3 \\
\hspace{1em} \textit{Java} & Java & 908 / 180 & 1:4 / 1:5 & 4955.7 $\pm$ 7583.8 & 428.1 $\pm$ 639.5 & 2.2 $\pm$ 3.5 \\
\hspace{1em} \textit{JavaScript} & JS & 1593 / 247 & 1:4 / 1:3 & 36723.8 $\pm$ 239449.4 & 715.8 $\pm$ 1074.2 & 2.0 $\pm$ 3.3 \\
\hspace{1em} \textit{Kotlin} & Kotlin & 5 / 6 & 0:1 / 1:5 & 7722.7 $\pm$ 9962.7 & 719.7 $\pm$ 940.8 & 1.6 $\pm$ 1.0 \\
\hspace{1em} \textit{Objective-C} & Obj-C & 19 / 0 & 1:5 / 0:0 & 8185.6 $\pm$ 4269.1 & 776.6 $\pm$ 403.6 & 1.1 $\pm$ 0.2 \\
\hspace{1em} \textit{PHP} & PHP & 4452 / 946 & 1:4 / 1:4 & 7943.1 $\pm$ 11725.2 & 627.5 $\pm$ 858.8 & 2.1 $\pm$ 5.6 \\
\hspace{1em} \textit{Python} & Python & 1032 / 180 & 1:4 / 1:4 & 8832.0 $\pm$ 11079.9 & 735.7 $\pm$ 864.9 & 1.9 $\pm$ 2.4 \\
\hspace{1em} \textit{Ruby} & Ruby & 409 / 75 & 1:4 / 1:3 & 5214.6 $\pm$ 7590.1 & 489.8 $\pm$ 689.7 & 1.7 $\pm$ 2.5 \\
\hspace{1em} \textit{Rust} & Rust & 79 / 12 & 1:9 / 1:11 & 7414.9 $\pm$ 6835.6 & 654.6 $\pm$ 603.9 & 1.8 $\pm$ 2.2 \\
\hspace{1em} \textit{Swift} & Swift & 23 / 6 & 1:2 / 1:2 & 8163.0 $\pm$ 12955.4 & 688.9 $\pm$ 986.4 & 3.3 $\pm$ 9.5 \\
\hspace{1em} \textit{TypeScript} & TS & 315 / 64 & 1:5 / 1:3 & 3606.0 $\pm$ 5916.6 & 364.2 $\pm$ 586.1 & 1.8 $\pm$ 3.0 \\
\hspace{1em} \textit{unknown} & N/A & 3148 / 510 & 1:4 / 1:4 & 13077.9 $\pm$ 25644.2 & 812.6 $\pm$ 1162.8 & 1.3 $\pm$ 1.9 \\
\cmidrule(l){2-7}
\hspace{1em} \textbf{Total} & Mix & 19161 / 2775 & 1:4 / 1:4 & 13660.4 $\pm$ 71581.8 & 894.6 $\pm$ 1145.6 & 2.2 $\pm$ 5.5 \\

\bottomrule
\end{tabular}%
}
\end{table*}
\vspace{-1em}

\subsection{Evaluation Metrics}
\label{sec:metrics}

To provide a comprehensive assessment of SAGE, we report standard classification metrics including Precision, Recall, and F1-Score. However, given the extreme class imbalance inherent in vulnerability detection datasets (e.g., BigVul contains only 6\% vulnerable samples~\cite{bigvul}, and PrimeVul 2.2\% ~\cite{primevul}), standard accuracy can be misleading, as a trivial model predicting the majority class would achieve high performance. To address this, we prioritize two robust metrics: \textbf{Balanced Accuracy (B.Acc)}~\cite{balanced_acc} and the \textbf{Matthews Correlation Coefficient (MCC)}~\cite{MCC}, which is defined as the arithmetic mean of sensitivity and specificity, ensuring equal weight for both minority and majority classes.

Balanced Accuracy (B.Acc) calculates the arithmetic mean of sensitivity (Recall) and specificity, preventing the evaluation from being dominated by the safe class (majority):
\begin{equation}
\text{Balanced Acc} = \frac{1}{2} \left( \frac{TP}{TP + FN} + \frac{TN}{TN + FP} \right),
\end{equation}
where $TP$, $TN$, $FP$, and $FN$ denote True Positives, True Negatives, False Positives, and False Negatives, respectively.

Furthermore, we employ the Matthews Correlation Coefficient (MCC), widely regarded as the most informative single metric for imbalanced binary classification. As demonstrated by Chicco and Jurman ~\cite{compare}, MCC is more informative than F1-score and accuracy in evaluating binary classification problems on imbalanced datasets, as it considers all four confusion matrix categories symmetrically:
\begin{equation}
\text{MCC} = \frac{TP \times TN - FP \times FN}{\sqrt{(TP+FP)(TP+FN)(TN+FP)(TN+FN)}},
\end{equation}
MCC values range from $-1$ to $+1$, where $+1$ indicates perfect prediction, $0$ indicates random guessing, and $-1$ indicates total disagreement.

\subsection{Implementation Details}
\label{sec:implementation}

All experiments were conducted on NVIDIA GeForce RTX 3090 GPUs. We implemented SAGE using PyTorch and the HuggingFace ecosystem.

\textbf{SAGE Configuration.} For the SAE component, we set the expansion factor $F=16$, mapping the hidden dimension to an overcomplete space of $F \times d_{model}$. The model was trained for 10 epochs with a batch size of 256 and a learning rate of $1\mathrm{e}{-4}$. The coefficients for the composite loss function (Eq. \ref{eq:loss_total}) were set to balance the competing objectives: sparsity penalty weight $\lambda=0.5$, and classification weight $\gamma=1.0$. We utilized an auxiliary $L_0$ coefficient of 1.0 to govern the sparsity thresholding. Unless otherwise stated, all task-conditional SAEs and their downstream classifiers are trained separately for each dataset-backbone pair; no sparse representation is shared across datasets.

\textbf{Overlength Handling.}
PreciseBugs contains many long files that may exceed the context limits of some backbones. To ensure a fair and reproducible comparison, we apply an explicit span-centered windowing strategy to overlong samples. Specifically, for each buggy or fixed file, we construct a fixed 4096-token input window centered on the annotated buggy or fixing span whenever possible. If the annotated span itself exceeds this budget, we anchor the window at the span start and keep the earliest 4096 tokens covering that region. The same preprocessing policy is applied consistently to SAGE and to all compared methods that operate on raw code inputs, so that performance differences are not confounded by inconsistent truncation rules. We will release this preprocessing step together with the data pipeline to facilitate reproduction.

\textbf{Implementation Details.} We adhered to official optimal settings, adapting for hardware constraints (all models trained for 3 epochs):
\begin{itemize}
\item \textbf{LineVul}~\cite{linevul}: Trained with learning rate (LR) $1\mathrm{e}{-5}$, block size 512, batch size 2, and gradient clipping norm 1.0.
\item \textbf{MSIVD}~\cite{yang2024securityvulnerabilitydetectionmultitask}: Utilized 4-bit QLoRA ($r=96, \alpha=32$, dropout 0.05), LR $1\mathrm{e}{-4}$, and sequence length 4096.
\item \textbf{VulLLM}~\cite{du-etal-2024-generalization}: Employed LoRA ($r=16, \alpha=32$, dropout 0.1) with cosine scheduler (0.1 warmup). Set LR to $2\mathrm{e}{-4}$, per-device batch size 2 with 8 gradient accumulation steps (FP16).
\item \textbf{ReVD}~\cite{wen2025boostingvulnerabilitydetectionllms}: Trained using IPO objective ($\beta=0.1$) and LoRA on all linear layers. Configured with LR $5\mathrm{e}{-6}$, batch size 1 with 8 accumulation steps (BF16).
\item \textbf{SteerLLM}~\cite{Li2025Steering}: Followed original paired-data configs. For PreciseBugs (unpaired), we derived steering vectors by subtracting mean activations of non-vulnerable from vulnerable instances.
\end{itemize}

\section{Experimental Results}
\subsection{RQ1: Overall Effectiveness}
\label{sec:rq1}

We evaluated SAGE against 11 baselines on BigVul~\cite{bigvul}, PrimeVul~\cite{primevul}, and the zero-day benchmark PreciseBugs~\cite{precisebugs} across four model families. Due to extreme class imbalance (e.g., 2.2\% positive rate in PrimeVul), we prioritize Matthews Correlation Coefficient (MCC) over Balanced Accuracy (B.Acc) and F1, as MCC provides a truthful representation of discriminative capability by symmetrically penalizing false positives and negatives~\cite{compare}.

\begin{table*}[t]
\centering
\caption{Comprehensive effectiveness comparison across different paradigms and backbones. The Prompt, SFT, RL, Activation, and Ours categories are evaluated across four distinct model families: CodeLlama-7B~\cite{codellama}, Mistral-7B~\cite{mistral7b}, Llama-3.1-8B~\cite{llama31}, and Qwen2.5-Coder-7B~\cite{qwen25}.}
\label{tab:main_results}
\vspace{-1em}

\renewcommand{\arraystretch}{0.95} % 稍微压缩行高以容纳30行数据
\setlength{\tabcolsep}{1.6pt} 

\resizebox{\textwidth}{!}{%
    \begin{threeparttable}
        \begin{tabular}{llccccccccccccccc}
        \toprule
        \multirow{2}{*}{\textbf{Category}} & \multirow{2}{*}{\textbf{Method (Backbone)}} & \multicolumn{5}{c}{\textbf{BigVul}~\cite{bigvul}} & \multicolumn{5}{c}{\textbf{PrimeVul}~\cite{primevul}} & \multicolumn{5}{c}{\textbf{PreciseBugs}~\cite{precisebugs}} \\
        \cmidrule(lr){3-7} \cmidrule(lr){8-12} \cmidrule(lr){13-17}
        & & Prec.~$\uparrow$ & Recall~$\uparrow$ & F1~$\uparrow$ & B.Acc~$\uparrow$ & \textbf{MCC}~$\uparrow$ 
          & Prec.~$\uparrow$ & Recall~$\uparrow$ & F1~$\uparrow$ & B.Acc~$\uparrow$ & \textbf{MCC}~$\uparrow$ 
          & Prec.~$\uparrow$ & Recall~$\uparrow$ & F1~$\uparrow$ & B.Acc~$\uparrow$ & \textbf{MCC}~$\uparrow$ \\
        \midrule

        % --- 1. Graph (2 rows) ---
        \multirow{2}{*}{\textbf{Graph}} 
         & ReVeal~\cite{deeplearning} & 6.68 & 45.94 & 11.67 & 54.01 & 0.0379 & 3.84 & 78.66 & 7.33 & 55.09 & 0.0396 & -- & -- & -- & -- & -- \\
         & IVDetect~\cite{Li_2021} & 20.13 & 49.62 & 28.65 & 57.32 & 0.1084 & 13.26 & 25.14 & 17.37 & 58.67 & 0.1287 & -- & -- & -- & -- & -- \\
        \midrule

        % --- 2. Prompt (4 rows) ---
        \multirow{4}{*}{\textbf{Prompt}} 
         & CodeLlama-7B~\cite{codellama} & 7.23 & 9.86 & 8.34 & 51.18 & 0.0204 & 5.12 & 22.59 & 8.34 & 56.55 & 0.0649 & 18.23 & 10.65 & 13.44 & 49.55 & -0.0113 \\
         & Mistral-7B~\cite{mistral7b} & 9.69 & 52.42 & 16.35 & 61.74 & 0.1174 & 3.20 & 48.09 & 6.01 & 57.59 & 0.0474 & 19.58 & 63.76 & 29.96 & 50.20 & 0.0033 \\
         & Llama-3.1-8B~\cite{llama31} & 5.39 & 95.73 & 10.21 & 48.11 & -0.1041 & 1.98 & 87.61 & 3.86 & 44.56 & -0.1222 & 19.44 & 98.15 & 32.46 & 49.88 & -0.0072 \\
         & Qwen2.5-Coder-7B~\cite{qwen25} & 6.28 & 75.83 & 11.59 & 54.38 & 0.0430 & 2.28 & 79.05 & 4.43 & 51.10 & 0.0077 & 18.71 & 57.84 & 28.27 & 48.52 & -0.0240 \\
        \midrule

        % --- 3. SFT (12 rows: 3 methods * 4 models) ---
        \multirow{12}{*}{\textbf{SFT}} 
         & LineVul (CodeLlama)~\cite{linevul} & 10.87 & 90.26 & 19.40 & 72.02 & 0.2072 & 4.50 & 84.82 & 8.54 & 69.54 & 0.1218 & 12.50 & 1.03 & 1.90 & 49.76 & -0.0153 \\
         & LineVul (Mistral) & 12.60 & 72.50 & 21.47 & 70.54 & 0.2043 & 4.11 & 88.70 & 7.85 & 68.07 & 0.1126 & 20.00 & 1.06 & 2.01 & 50.02 & 0.0013 \\
         & LineVul (Llama-3.1) & 13.30 & 93.60 & 23.29 & 77.75 & 0.2651 & 4.66 & 78.58 & 8.80 & 68.88 & 0.1190 & 14.29 & 1.03 & 1.92 & 49.87 & -0.0089 \\
         & LineVul (Qwen2.5) & 10.74 & 75.29 & 18.80 & 68.11 & 0.1731 & 5.06 & 79.09 & 9.51 & 70.69 & 0.1321 & 7.69 & 1.03 & 1.82 & 49.22 & -0.0392 \\
         \cmidrule(lr){2-17} % 分隔线区分不同 SFT 方法
         & VulLLM (CodeLlama)~\cite{du-etal-2024-generalization} & 8.14 & 86.73 & 14.88 & 64.36 & 0.1345 & 3.25 & 91.62 & 6.29 & 64.97 & 0.0909 & 18.43 & 78.19 & 29.83 & 47.22 & -0.0583 \\
         & VulLLM (Mistral) & 22.11 & 90.81 & 35.57 & 85.93 & 0.3925 & 2.43 & 98.36 & 4.74 & 54.44 & 0.0430 & 19.33 & 95.56 & 32.16 & 49.53 & -0.0199 \\
         & VulLLM (Llama-3.1) & 5.63 & \textbf{100.00} & 10.66 & 50.34 & 0.0196 & 2.31 & \textbf{99.82} & 4.51 & 52.07 & 0.0303 & 19.52 & \textbf{99.82} & 32.66 & 50.11 & 0.0144 \\
         & VulLLM (Qwen2.5) & 9.45 & 30.90 & 14.48 & 56.68 & 0.0794 & 2.50 & 32.24 & 4.64 & 51.87 & 0.0122 & 19.52 & 98.15 & 32.57 & 50.10 & 0.0059 \\
         \cmidrule(lr){2-17}
         & MSIVD (CodeLlama)~\cite{yang2024securityvulnerabilitydetectionmultitask} & 13.45 & 1.51 & 2.73 & 50.47 & 0.0272 & 13.45 & 2.91 & 4.79 & 51.24 & 0.0528 & 19.38 & 79.59 & 31.17 & 48.41 & -0.0332 \\
         & MSIVD (Mistral) & 10.15 & 75.36 & 17.89 & 67.92 & 0.1671 & 6.35 & 43.72 & 11.10 & 64.56 & 0.1192 & 19.14 & 58.06 & 28.78 & 48.36 & -0.0268 \\
         & MSIVD (Llama-3.1) & 5.51 & 98.39 & 10.44 & 49.26 & -0.0738 & 2.11 & 94.90 & 4.14 & 47.70 & -0.0874 & 19.49 & 98.89 & 32.56 & 50.02 & 0.0016 \\
         & MSIVD (Qwen2.5) & 76.92 & 0.95 & 1.87 & 50.47 & 0.0832 & 6.39 & 63.39 & 11.61 & 71.18 & 0.1506 & 20.00 & 0.44 & 0.86 & 50.01 & 0.0009\\
        \midrule

        % --- 4. RL (4 rows) ---
        \multirow{4}{*}{\textbf{RL}} 
         & ReVD (CodeLlama)~\cite{wen2025boostingvulnerabilitydetectionllms} & 18.80 & 93.65 & 31.31 & 84.84 & 0.3572 & 5.86 & 40.26 & 10.24 & 62.81 & 0.1050 & 19.44 & 92.05 & 32.10 & 49.85 & -0.0044 \\
         & ReVD (Mistral) & 66.14 & 71.09 & 68.52 & 84.47 & 0.6664 & 3.55 & 42.99 & 6.55 & 58.25 & 0.0548 & 19.58 & 97.41 & 32.61 & 50.29 & 0.0144\\
         & ReVD (Llama-3.1) & 7.72 & 93.46 & 14.26 & 63.63 & 0.1340 & 3.66 & 70.86 & 6.97 & 64.33 & 0.0852 & 19.34 & 97.04 & 32.26 & 49.55 & -0.0241 \\
         & ReVD (Qwen2.5) & 33.03 & 6.82 & 11.31 & 53.00 & 0.1290 & 5.49 & 4.19 & 4.75 & 51.28 & 0.0292 & 19.57 & 97.60 & 32.60 & 50.25 & 0.0121 \\
        \midrule

        % --- 5. Activation (4 rows) ---
        \multirow{4}{*}{\textbf{Activation}} 
         & SteerLLM (CodeLlama)~\cite{Li2025Steering} & 10.68 & 18.77 & 13.61 & 54.73 & 0.0730 & 3.24 & 53.19 & 6.11 & 58.62 & 0.0527 & 17.16 & 57.87 & 26.47 & 45.13 & -0.0813 \\
         & SteerLLM (Mistral) & 11.02 & 41.52 & 17.42 & 60.83 & 0.1220 & 3.33 & 38.80 & 6.13 & 56.63 & 0.0446 & 17.96 & 40.56 & 24.89 & 47.86 & -0.0341 \\
         & SteerLLM (Llama-3.1) & 5.41 & 96.11 & 10.25 & 48.32 & -0.0914 & 2.03 & 89.25 & 3.98 & 45.94 & -0.0722 & 20.54 & 63.49 & 31.03 & 50.22 & 0.0036 \\
         & SteerLLM (Qwen2.5) & 6.54 & 81.33 & 12.11 & 56.24 & 0.0623 & 2.08 & 83.61 & 4.06 & 47.24 & -0.0260 & 20.82 & 44.80 & 28.43 & 50.53 & 0.0087 \\
        \midrule
        \multirow{4}{*}{\textbf{Agent}} 
         & ConColl (CodeLlama)~\cite{concoll} & 5.22 & 91.00 & 9.87 & 46.55 & -0.1017 & 1.76 & 75.77 & 3.44 & 39.91 & -0.1432 & 19.12 & 80.40 & 30.89 & 49.06 & -0.0194 \\
         & ConColl (Mistral) & 11.35 & 35.17 & 17.16 & 59.44 & 0.1146 & 4.00 & 47.72 & 7.38 & 60.88 & 0.0726 & 19.40 & 68.20 & 30.21 & 49.82 & -0.0030 \\
         & ConColl (Llama-3.1) & 13.79 & 15.55 & 14.62 & 54.89 & 0.0926 & 14.44 & 27.66 & 18.98 & 47.31 & -0.0431 & 19.44 & 33.43 & 24.58 & 49.96 & -0.0007 \\
         & ConColl (Qwen2.5) & 17.50 & 1.33 & 2.47 & 50.48 & 0.0338 & 7.80 & 2.00 & 3.19 & 50.73 & 0.0287 & 20.92 & 4.22 & 7.02 & 50.18 & 0.0073 \\
        \midrule
        % --- Frontier ---
\multirow{3}{*}{\textbf{Frontier}} 
 & GPT-5.2-Codex~\cite{gpt52} & 10.15 & 52.38 & 17.00 & 62.52 & 0.1269 & 3.64 & 56.83 & 6.85 & 61.40 & 0.0909 & 19.11 & 46.38 & 27.07 & 49.54 & -0.0072 \\
 & Gemini 3 Pro~\cite{gemini3} & 8.45 & 71.28 & 15.11 & 62.77 & 0.1175 & 3.27 & 84.34 & 6.30 & 63.92 & 0.0828 & 18.87 & 60.48 & 28.76 & 48.90 & -0.0179 \\
 & Claude Opus 4.5~\cite{claude45} & 18.93 & 30.48 & 23.36 & 61.39 & 0.1829 & 4.86 & 85.45 & 9.19 & \textbf{73.73} & 0.1433 & 20.99 & 36.18 & 26.57 & 51.68 & 0.0282 \\
\midrule
        
        % --- 6. Ours (4 rows) ---
        \multirow{4}{*}{\textbf{Ours}} 
         & \textbf{SAGE} (CodeLlama) & 82.02\inc{307} & 77.82\dec{17} & \textbf{79.86}\inc{155} & \textbf{88.40}\inc{4} & \textbf{0.7874}\inc{120} & 17.87\inc{33} & 32.42\dec{65} & 23.04\inc{27} & 64.52\dec{12} & \textbf{0.2375}\inc{66} & 23.49\inc{12} & 67.70\dec{26} & 34.88\inc{9} & 54.79\inc{6} & 0.1050\inc{272} \\
         & \textbf{SAGE} (Mistral) & 77.22\inc{17} & 70.05\dec{23} & 73.46\inc{7} & 84.41\dec{0.07} & 0.7206\inc{8} & 17.70\inc{33} & 31.33\dec{68} & 22.62\inc{25} & 64.01\dec{13} & 0.2125\inc{48} & 22.55\inc{7} & 84.63\dec{13} & 35.61\inc{9} & 54.01\inc{5} & 0.1104\inc{291} \\
         & \textbf{SAGE} (Llama-3.1) & 84.70\inc{321} & 68.72\dec{31} & 75.88\inc{165} & 83.99\inc{8} & 0.7506\inc{183} & 17.78\inc{23} & 32.06\dec{68} & 22.87\inc{20} & 64.35\dec{13} & 0.2157\inc{51} & \textbf{24.91}\inc{19} & 69.60\dec{30} & \textbf{36.69}\inc{12} & \textbf{57.15}\inc{11} & \textbf{0.1178}\inc{318} \\
         & \textbf{SAGE} (Qwen2.5) & \textbf{86.97}\inc{13} & 69.57\dec{14} & 77.30\inc{170} & 84.48\inc{24} & 0.7664\inc{319} & \textbf{19.16}\inc{44} & 31.69\dec{63} & \textbf{23.88}\inc{37} & 64.33\dec{13} & 0.2245\inc{57} & 22.90\inc{9} & 73.06\dec{26} & 34.87\inc{7} & 54.12\inc{5} & 0.0986\inc{250} \\
        \bottomrule
        \end{tabular}
        
        \begin{tablenotes}[flushleft]
    \footnotesize
    \item \textit{Note:} To ensure a rigorous and fair comparison of architectural effectiveness, all open-source baselines and SAGE variants are evaluated using backbones within the \textbf{7B--8B parameter range}: \textbf{CodeLlama-7B}, \textbf{Mistral-7B}, \textbf{Qwen2.5-Coder-7B}, and \textbf{Llama-3.1-8B}. 
    Red (e.g., \inc{295}) and green (e.g., \dec{17}) percentages denote relative improvement and reduction, respectively, compared to the \textbf{best performance} among methods using the same backbone, Graph-based approaches, and commercial models.
    Metrics marked with $\uparrow$ indicate higher is better. 
    The best results across all methods are highlighted in \textbf{bold}.
    \textbf{Results for Graph-based baselines (ReVeal, IVDetect) on PreciseBugs are omitted due to toolchain limitations: their reliance on the Joern C/C++ parser precludes execution on the multi-language composition of this dataset.}
\end{tablenotes}
    \end{threeparttable}%
}
\vspace{-1em}
\end{table*}

 \textbf{Superiority and Generalization.} 
As shown in Table~\ref{tab:main_results}, SAGE consistently achieves State-of-the-Art (SOTA) performance. While structural methods (e.g., ReVeal) struggle with parsing failures and standard SFT/Prompting strategies performance degrade on the zero-day PreciseBugs benchmark (yielding near-zero or negative MCCs), SAGE exhibits improved generalization. Specifically, \textbf{SAGE (CodeLlama)} sets new records on BigVul (\textbf{0.7874} MCC) and PrimeVul (\textbf{0.2375} MCC). On PreciseBugs, \textbf{SAGE (Llama-3.1)} attains the highest MCC of \textbf{0.1178}, outperforming the proprietary frontier model Claude Opus 4.5~\cite{claude45}. 
Crucially, SAGE proves to be backbone-agnostic; for instance, it boosts Llama-3.1's MCC on BigVul from negative values (in Prompting/MSIVD) to a robust \textbf{0.7506}, validating that the Sparse Autoencoder successfully disentangles security features regardless of the underlying architecture.

 \textbf{The Limitation of Pre-training Knowledge.}
Despite the improvements, the absolute performance decreases on unseen datasets like PrimeVul and PreciseBugs. We attribute this to a lack of prior knowledge in the base model.
SAGE is designed to recover hidden vulnerability signals that already exist within the model's representations. However, it cannot compensate for the total absence of knowledge regarding new attack vectors that were not present in the pre-training data.
This result suggests that even with our enhancements, the model cannot effectively reason about completely new types of attacks without continuous learning.

 \textbf{Analysis of Baseline Failure.}
Table~\ref{tab:main_results} reveals a issue with the baseline models.
Methods like VulLLM and MSIVD exhibit near-perfect Recall ($>$98\%) but extremely low Precision ($<$5\%) on unseen data. This indicates that these models tend to classify almost everything as vulnerable when they handle noisy real-world data. They lower their decision thresholds to avoid missing any potential defects. While this strategy inflates Balanced Accuracy, the low MCC scores demonstrate that they fail to effectively distinguish between safe and vulnerable code. In contrast, SAGE acts as a precise detector.
The sparsity constraint in our SAE serves as a strict filter. It removes ambiguous signals and reduces false alarms. Although this approach results in lower Recall, it ensures that the detections are reliable, leading to higher Precision and MCC.
\vspace{-0.7em}
\begin{boxK}
\small \faIcon{pencil-alt} \textbf{Finding 1 (Generalization \& Reliability):} Existing paradigms exhibit a "High-Recall, Low-Precision" pattern on zero-day benchmarks, degrading to random guessing. SAGE overcomes this limitation by actively filtering semantic noise, sacrificing trivial recall to achieve high-precision, robust generalization across unseen distributions.
\end{boxK}
\vspace{-0.3em}

\begin{figure*}[h]
\centering
\includegraphics[width=1\linewidth]{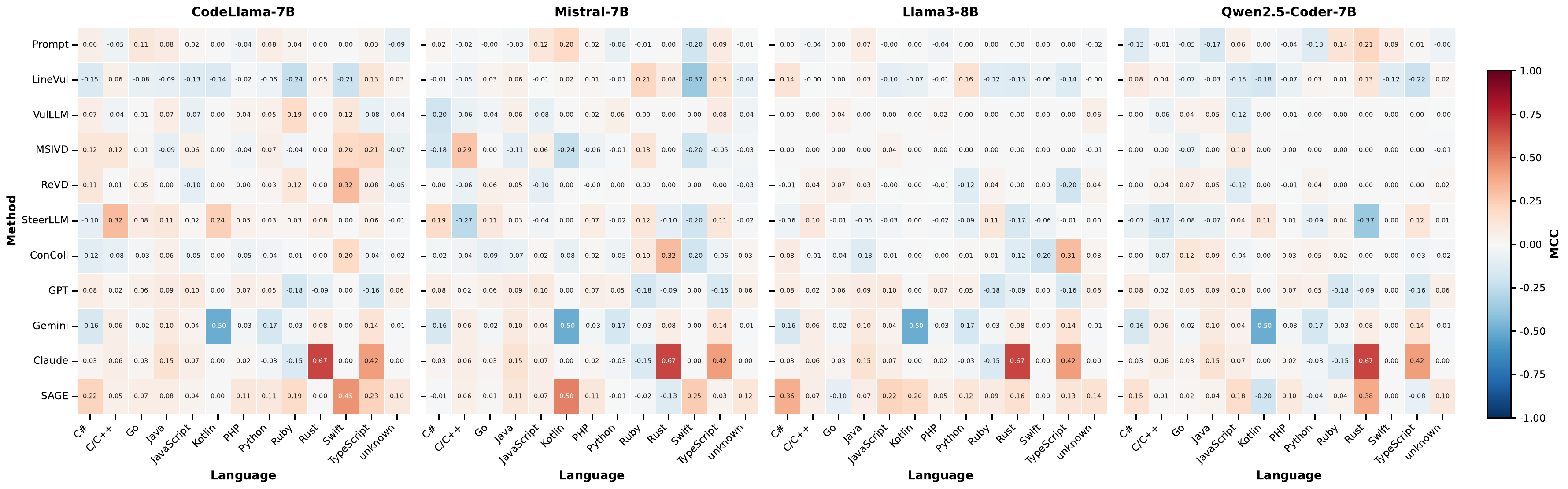}
\caption{Fine-grained Language-wise Generalization on the zero-day PreciseBugs benchmark. The heatmaps visualize MCC scores across 13 programming languages for different methods and backbones. Red (warm) indicates positive correlation, while blue (cold) indicates negative correlation.}
\label{fig:heatmap}
\vspace{-0.7em}
\end{figure*}

 \textbf{Robustness Across Different Languages.}
Fig.~\ref{fig:heatmap} exposes a second consistent limitations in baseline models.
Traditional methods and even frontier models perform poorly on less common programming languages.
A notable example is \textbf{Gemini 3 Pro}, which suffers a performance drop to \textbf{-0.50 MCC} on Kotlin.
This suggests that these models rely heavily on the surface-level syntax of the languages they were trained on.
They fail to identify defects when the syntax changes.
In contrast, SAGE shows positive transfer across the 13-language PreciseBugs split. However, languages with very small test sets (e.g., Kotlin, Swift, and Rust) should be interpreted as qualitative case studies rather than stable performance estimates.
By using sparse projection, SAGE effectively separates the logic of the defect from the syntactic style of the code.
This capability allows a 7B model to detect defects in languages where trillion-parameter generalist models fail due to unfamiliar syntax.

\vspace{-0.3em}
\begin{boxK}
\small \faIcon{pencil-alt} \textbf{Finding 2 (Generalization to Long-tail Syntax):} Strong performance on mainstream languages may not generalize to diverse syntax patterns. Most methods struggle to maintain accuracy on long-tail languages like Kotlin. SAGE overcomes this limitation. It achieves consistent detection rates across all tested languages. This proves that the method successfully learns universal defect features instead of memorizing specific language syntax.
\end{boxK}
\vspace{-1em}

\subsection{RQ2: Mechanistic Verification}
\label{sec:rq2}

 \textbf{Analysis of Signal Loss.}
To rigorously quantify the phenomenon of signal loss, we employ a dual-metric approach that analyzes the evolution of vulnerability features.
This method allows us to verify two complementary aspects: the linear separability of the signal and its relative dominance against the background semantics.
First, we evaluate discriminability using the Projected Signal-to-Noise Ratio (SNR).
Second, we measure the dominance of the signal using the Magnitude Ratio ($\rho_l$), which compares the strength of vulnerability signals ($v_l$) against global semantics ($s_l$).
Guided by the Linear Representation Hypothesis~\cite{Park2023}, we first define the class centroids at layer $l$ as
$\mu_{\mathrm{vul}}^{(l)} = \mathrm{Mean}(H_{\mathrm{vuln}}^{(l)})$ and
$\mu_{\mathrm{safe}}^{(l)} = \mathrm{Mean}(H_{\mathrm{safe}}^{(l)})$,
and use their difference to construct the vulnerability direction
$v_l = \mu_{\mathrm{vul}}^{(l)} - \mu_{\mathrm{safe}}^{(l)}$.
We then normalize this direction as $\hat{v}_l = v_l / \|v_l\|_2$, and project each activation $h_l(x)$ onto it via
$p_l(x) = h_l(x)^\top \hat{v}_l$.
Accordingly, $\mu'_{\mathrm{vul}}$, $\mu'_{\mathrm{safe}}$ and $\mathrm{Var}'_{\mathrm{vul}}$, $\mathrm{Var}'_{\mathrm{safe}}$ denote the class-wise means and population variances of these scalar projections over vulnerable and safe samples, respectively.
We formulate these metrics at layer $l$ as follows:
\vspace{-0.5em}
\begin{equation}
\text{SNR}_l = \frac{(\mu'_{\mathrm{vul}} - \mu'_{\mathrm{safe}})^2}{\mathrm{Var}'_{\mathrm{vul}} + \mathrm{Var}'_{\mathrm{safe}} + \epsilon}, \quad \quad
\rho_l = \frac{\|v_l\|_2}{\|s_l\|_2}.
\label{eq:snr_rho}
\end{equation}

\begin{figure*}[h]
\centering
\includegraphics[width=0.85\linewidth]{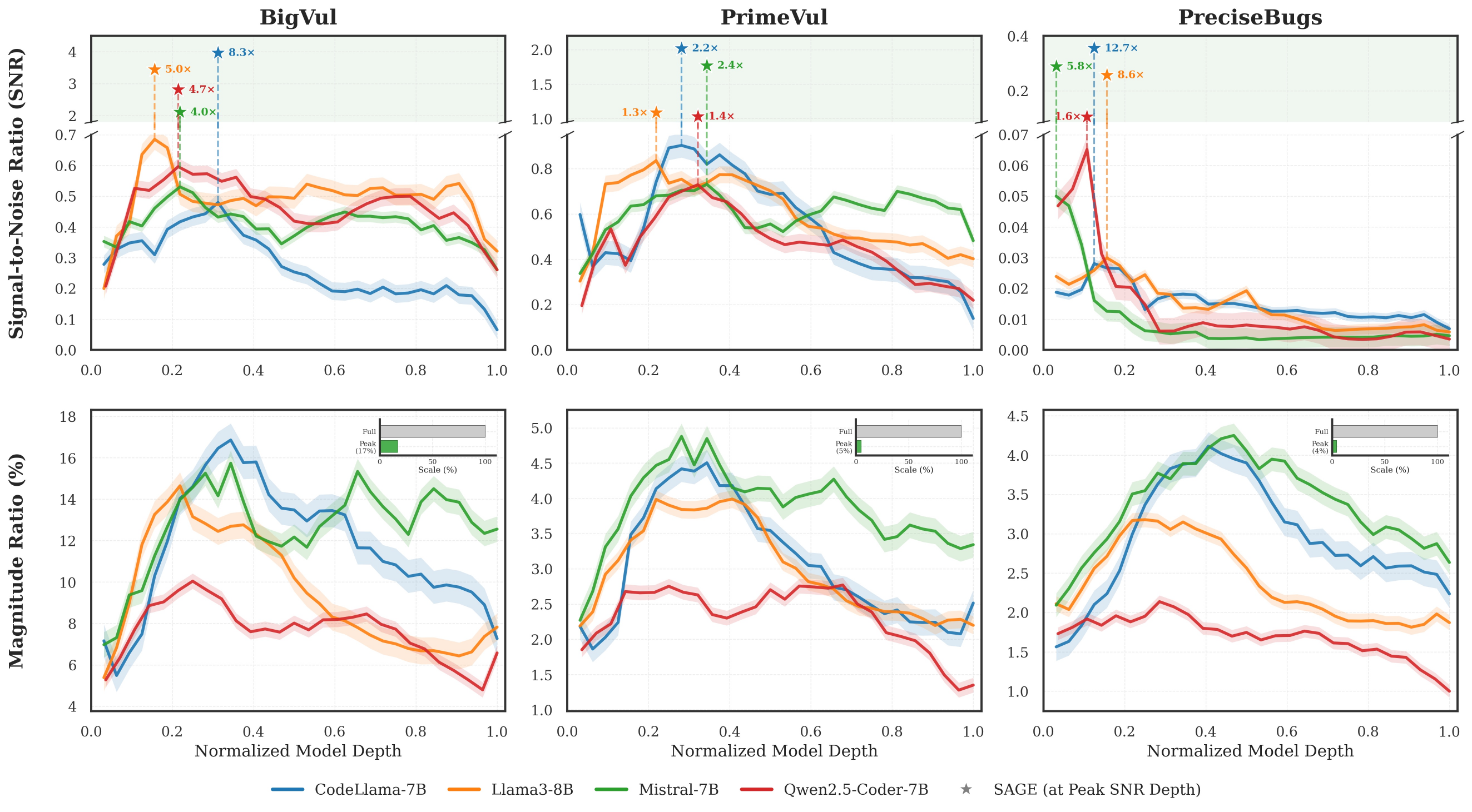}
\vspace{-1em}
\caption{Layer-wise evolution of SNR and Feature Magnitude Ratio. The signal follows an "Inverted-U" trajectory, decaying in deep layers. SAGE (stars) successfully recovers the peak signal.}
\label{fig:snr_magnitude}
\vspace{-1.5em}
\end{figure*}
 \textbf{The Reality of Signal Submersion.}
Fig. \ref{fig:snr_magnitude} visualizes these metrics across four model architectures.
The SNR results reveal an inverted-U trajectory where vulnerability signals peak in intermediate layers before collapsing in the deeper layers.
This trend implies that standard models perform inference using weakened signals in the final layer.
The Magnitude Ratio provides the explanation for this degeneration.

As shown in the bottom of Fig. \ref{fig:snr_magnitude}, the vulnerability activations constitute less than 5\% of the total energy ($\rho_l < 0.05$).
This low ratio indicates that the subtle vulnerability features are substantially dominated by the dominant global semantics.
SAGE effectively addresses this imbalance.
On the PreciseBugs benchmark using CodeLlama-7B, SAGE amplifies the SNR by \textbf{12.7$\times$} compared to the baseline peak.
This significant increase validates that our method successfully recovers submerged signals by boosting their relative magnitude.

\begin{figure*}[h]
\centering
\includegraphics[width=0.8 \linewidth]{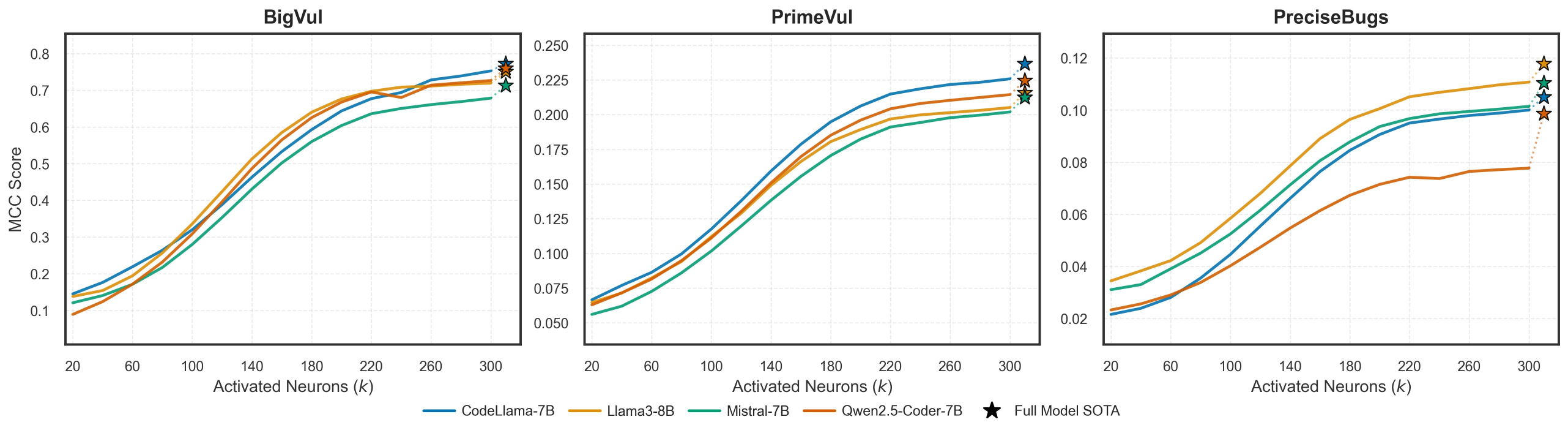}
\vspace{-0.5em}
\caption{Top-$K$ Feature Attribution. Performance saturates with fewer than 250 atomic features, confirming the compactness of the extracted vulnerability basis.}
\label{fig:topk_sparsity}
\vspace{-1em}
\end{figure*}

 \textbf{Analysis of Representational Sparsity.}
We further investigate the efficiency of the learned sparse representation by restricting inference to the top-$K$ vulnerability-relevant neurons.
In Fig.~\ref{fig:topk_sparsity}, the top-$K$ selection is performed globally at the dataset level rather than independently for each sample.
Specifically, we first identify neurons that are clearly activated on vulnerable samples but not on safe samples, rank them by their average activation magnitude on vulnerable samples, and retain only the top $K$ such features during inference.
As illustrated in Fig.~\ref{fig:topk_sparsity}, the performance exhibits a rapid increase followed by a stable plateau.
The model recovers 95\% of its peak detection capability using only about 220 active neurons.
This small subset constitutes less than 1\% of the total feature space.
This observation indicates that vulnerability-relevant information is highly concentrated in a small number of sparse features.
It further suggests that SAGE learns a precise and efficient representation, rather than relying on a large number of diffuse parameters to memorize the data.

\vspace{-0.5em}
\begin{boxK}
\small \faIcon{pencil-alt} \textbf{Finding 3 (Mechanism of Improvement):} SAGE mitigates signal loss by refining the feature representation. Standard models often suppress vulnerability features to less than 5\% of the total strength in deep layers. The sparse structure of SAGE corrects this issue. It amplifies the SNR by 12.7$\times$ and concentrates the critical information. Consequently, the complex decision logic is reduced to a concise set of approximately 250 essential neurons.    
\end{boxK}
\vspace{-0.5em}

\subsection{RQ3: Component Efficacy \& Stability}
\label{sec:rq3}
To determine whether SAGE's performance stems from its architectural design rather than incidental tuning, we conducted a rigorous ablation study and hyperparameter sensitivity analysis.

 \textbf{Analysis of Components and Architecture.}
 
The results in Table \ref{tab:ablation_study} clarify the contribution of each specific module.
Although the Linear Probe outperforms standard finetuning, its stagnation at a low Precision of approximately 25\% indicates that vulnerability signals remain deeply mixed within the complex feature space.
This limitation confirms that simple linear methods are insufficient to separate these entangled signals.

Standard aggregation strategies similarly exhibit high instability across different datasets.
Mean Pooling tends to dilute the specific vulnerability signals by averaging the features, while Max Pooling often focuses incorrectly on strong noise signals rather than subtle defects.
This failure is particularly evident on the PreciseBugs dataset where Max Pooling yields an MCC near 0.00 because it prioritizes high-energy noise over the actual vulnerability patterns.

In contrast, the unsupervised SAE functions as an effective noise filter and substantially improves Precision to 74.62\% on the BigVul dataset.
However, this method lacks specific semantic guidance and consequently suffers from a drop in Recall to 55.45\%.
Full SAGE resolves this trade-off by using classification gradients to refine the feature representation.
This approach effectively restores Recall to 77.82\% while maintaining a state-of-the-art Precision of 82.02\%.
Furthermore, SAGE achieves consistent gains across CodeLlama, Mistral, Llama-3.1, and Qwen2.5, demonstrating that the method learns universally transferable features rather than relying on specific model architectures.

\begin{table*}[h]
\centering
\caption{Ablation study of SAGE components. We compare \textbf{Full SAGE} with four variants: Vanilla Finetuning, Linear Probe, Unsupervised SAE, and Pooling strategies.}
\label{tab:ablation_study}
\vspace{-1em}

\renewcommand{\arraystretch}{0.93} % 稍微增加行高，提升可读性
\setlength{\tabcolsep}{2.5pt}      % 稍微调整列间距
\resizebox{\textwidth}{!}{%
    \begin{threeparttable}
        \begin{tabular}{llccccccccccccccc}
        \toprule
        \multirow{2}{*}{\textbf{Variant}} & \multirow{2}{*}{\textbf{Backbone}} & \multicolumn{5}{c}{\textbf{BigVul}~\cite{bigvul}} & \multicolumn{5}{c}{\textbf{PrimeVul}~\cite{primevul}} & \multicolumn{5}{c}{\textbf{PreciseBugs}~\cite{precisebugs}} \\
        \cmidrule(lr){3-7} \cmidrule(lr){8-12} \cmidrule(lr){13-17}
        & & Prec.~$\uparrow$ & Recall~$\uparrow$ & F1~$\uparrow$ & B.Acc~$\uparrow$ & \textbf{MCC}~$\uparrow$ 
          & Prec.~$\uparrow$ & Recall~$\uparrow$ & F1~$\uparrow$ & B.Acc~$\uparrow$ & \textbf{MCC}~$\uparrow$ 
          & Prec.~$\uparrow$ & Recall~$\uparrow$ & F1~$\uparrow$ & B.Acc~$\uparrow$ & \textbf{MCC}~$\uparrow$ \\
        \midrule

        % --- 1. Vanilla Finetuning (Baseline) ---
        \multirow{4}{*}{\shortstack[l]{\textbf{Vanilla}\\\textbf{Finetuning}}} 
          & CodeLlama-7B     & 10.87 & 90.26 & 19.40 & 72.02 & 0.2072 & 4.50 & 84.82 & 8.54 & 69.54 & 0.1218 & 12.50 & 1.03 & 1.90 & 49.76 & -0.0153 \\
          & Mistral-7B       & 12.60 & 72.50 & 21.47 & 70.54 & 0.2043 & 4.11 & \textbf{88.70} & 7.85 & 68.07 & 0.1126 & 20.00 & 1.06 & 2.01 & 50.02 & 0.0013 \\
          & Llama-3.1-8B     & 13.30 & \textbf{93.60} & 23.29 & 77.75 & 0.2651 & 4.66 & 78.58 & 8.80 & 68.88 & 0.1190 & 14.29 & 1.03 & 1.92 & 49.87 & -0.0089 \\
          & Qwen2.5-Coder-7B & 10.74 & 75.29 & 18.80 & 68.11 & 0.1731 & 5.06 & 79.09 & 9.51 & 70.69 & 0.1321 & 7.69 & 1.03 & 1.82 & 49.22 & -0.0392 \\
        \midrule
       
        % --- 2. Pan-Layer + Linear Probe ---
        \multirow{4}{*}{\shortstack[l]{\textbf{Linear}\\\textbf{Probe}}} 
          & CodeLlama-7B     & 25.28 & 76.87 & 38.05 & 81.71 & 0.3878 & 6.08 & 69.76 & 11.18 & 72.67 & 0.1532 & 15.38 & 2.13 & 3.74 & 49.88 & -0.0058 \\
          & Mistral-7B       & 24.73 & 68.63 & 36.35 & 78.12 & 0.3569 & 5.97 & 76.14 & 11.08 & 74.50 & 0.1602 & 15.56 & 3.19 & 5.29 & 49.50 & -0.0202 \\
          & Llama-3.1-8B     & 28.62 & 75.92 & 41.57 & 82.35 & 0.4183 & 6.01 & 76.87 & 11.16 & \textbf{74.83} & 0.1622 & 12.50 & 2.27 & 3.85 & 49.60 & -0.0174 \\
          & Qwen2.5-Coder-7B & 29.71 & 72.61 & 42.17 & 81.22 & 0.4176 & 6.16 & 73.95 & 11.38 & 74.22 & 0.1614 & 14.29 & 6.38 & 8.82 & 49.33 & -0.0190 \\
        \midrule

        % --- 3. SAGE-Unsupervised ---
        \multirow{4}{*}{\shortstack[l]{\textbf{Unsupervised}\\\textbf{(w/o $\mathcal{L}_{cls}$})}} 
          & CodeLlama-7B     & 74.62 & 55.45 & 63.62 & 77.17 & 0.6257 & 14.31 & 29.33 & 19.24 & 62.67 & 0.1791 & 20.94 & 98.10 & 34.51 & 50.23 & 0.0707 \\
          & Mistral-7B       & 73.18 & 51.47 & 60.43 & 75.18 & 0.5953 & 14.17 & 27.50 & 18.70 & 61.86 & 0.1721 & 21.27 & 94.47 & 34.72 & 51.15 & 0.0628 \\
          & Llama-3.1-8B     & 70.04 & 54.50 & 61.30 & 76.56 & 0.5983 & 14.14 & 29.87 & 19.19 & 62.88 & 0.1795 & 21.99 & 86.18 & 35.04 & 52.81 & 0.0697 \\
          & Qwen2.5-Coder-7B & 80.31 & 54.88 & 65.20 & 77.04 & 0.6481 & 15.26 & 28.05 & 19.77 & 62.26 & 0.1826 & 21.74 & 82.56 & 34.41 & 52.11 & 0.0597 \\
        \midrule

        % --- 4. Pooling (Split into Mean and Max) ---
        \multirow{4}{*}{\textbf{Pooling (Mean)}} 
          & CodeLlama-7B     & 59.43 & 72.61 & 65.36 & 84.83 & 0.6344 & 13.48 & 46.63 & 20.92 & 69.93 & 0.2205 & 18.37 & 38.30 & 24.83 & 51.98 & 0.0310 \\
          & Mistral-7B       & 57.27 & 64.93 & 60.86 & 81.03 & 0.5852 & 15.13 & 26.05 & 19.14 & 61.37 & 0.1747 & 17.07 & 97.87 & 29.07 & 50.97 & 0.0382 \\
          & Llama-3.1-8B     & 58.71 & 62.94 & 60.75 & 80.16 & 0.5838 & 10.97 & 47.91 & 17.85 & 69.55 & 0.1947 & 17.25 & 98.94 & 29.38 & 51.61 & 0.0634 \\
          & Qwen2.5-Coder-7B & 62.06 & 70.24 & 65.90 & 83.85 & 0.6389 & 12.88 & 32.79 & 18.50 & 63.88 & 0.1771 & 17.22 & 60.64 & 26.82 & 50.92 & 0.0140 \\
        \cmidrule(l){1-1} % 小分割线，区分Mean和Max，但保持连贯
        \multirow{4}{*}{\textbf{Pooling (Max)}} 
          & CodeLlama-7B     & 33.21 & 43.98 & 37.85 & 69.37 & 0.3399 & 15.50 & 23.31 & 18.62 & 60.22 & 0.1676 & 0.00 & 0.00 & 0.00 & 50.00 & 0.0000 \\
          & Mistral-7B       & 63.11 & 50.43 & 56.06 & 74.34 & 0.5414 & 14.70 & 28.78 & 19.46 & 62.50 & 0.1806 & 16.85 & 98.94 & 28.79 & 50.22 & 0.0139 \\
          & Llama-3.1-8B     & 53.68 & 52.51 & 53.09 & 74.91 & 0.5034 & 13.84 & 28.60 & 18.66 & 62.28 & 0.1730 & 16.79 & \textbf{100.00} & 28.75 & 50.00 & 0.0000 \\
          & Qwen2.5-Coder-7B & 58.01 & 42.56 & 49.10 & 70.37 & 0.4719 & 15.22 & 22.40 & 18.13 & 59.79 & 0.1623 & 19.79 & 81.91 & 31.88 & \textbf{57.48} & 0.0000 \\
        \midrule

        % --- 5. SAGE (Full) ---
        \multirow{4}{*}{\shortstack[l]{\textbf{Full}\\\textbf{SAGE}}} 
          & CodeLlama-7B     & 82.02 & 77.82 & \textbf{79.86} & \textbf{88.40} & \textbf{0.7874} & 17.87 & 32.42 & \textbf{23.04} & 64.52 & \textbf{0.2375} & 23.49 & 67.70 & 34.88 & 54.79 & 0.1050 \\
          & Mistral-7B       & 77.22 & 70.05 & 73.46 & 84.41 & 0.7206 & 17.70 & 31.33 & 22.62 & 64.01 & 0.2125 & 22.55 & 84.63 & 35.61 & 54.01 & 0.1104 \\
          & Llama-3.1-8B     & 84.70 & 68.72 & 75.88 & 83.99 & 0.7506 & 17.78 & 32.06 & 22.87 & 64.35 & 0.2157 & \textbf{24.91} & 69.60 & \textbf{36.69} & 57.15 & \textbf{0.1178} \\
          & Qwen2.5-Coder-7B & \textbf{86.97} & 69.57 & 77.30 & 84.48 & 0.7664 & \textbf{19.16} & 31.69 & 23.88 & 64.33 & 0.2245 & 22.90 & 73.06 & 34.87 & 54.12 & 0.0986 \\
        \bottomrule
        \end{tabular}
        
        \begin{tablenotes}[flushleft]
    \footnotesize
    \item \textit{Note:}
    \textbf{Linear Probe} results are derived from L2-regularized logistic regression on frozen representations. 
    The best results across all methods are highlighted in \textbf{bold}.
\end{tablenotes}
    \end{threeparttable}%
}
\vspace{-1em}
\end{table*}

% \vspace{-0.7em}
\begin{figure}[h]
    \centering
    \begin{subfigure}[b]{1\linewidth}
        \centering
        \includegraphics[width=\linewidth]{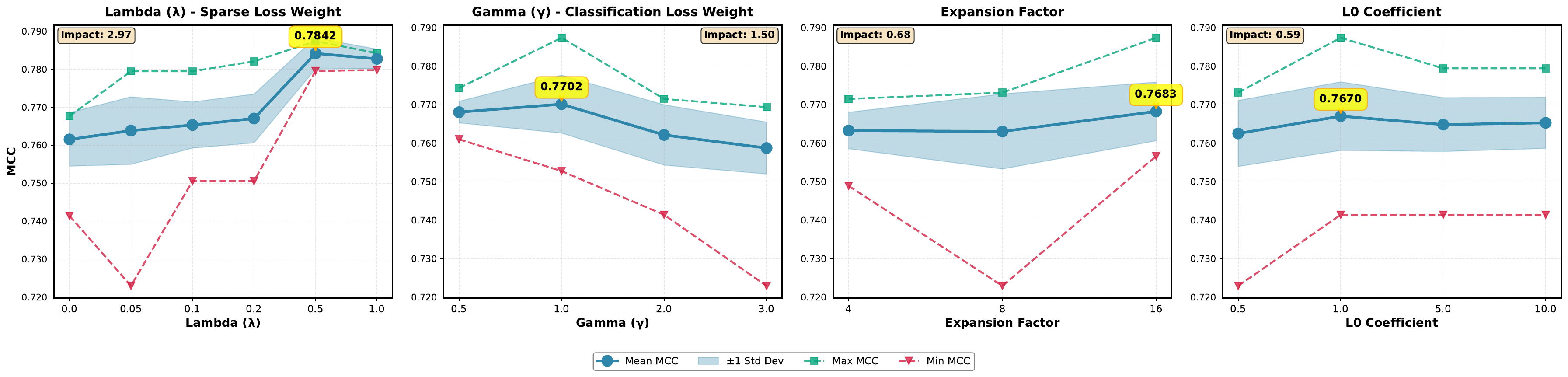}
        \caption{Impact of key hyperparameters ($\lambda$, $\gamma$, $L_0$ coeff, and Expansion Factor) on model performance at the optimal intervention depth.}
        \label{fig:hyperparam_sensitivity}
    \end{subfigure}
    \vspace{1em} 
    \begin{subfigure}[b]{1\linewidth}
        \centering
        \includegraphics[width=\linewidth]{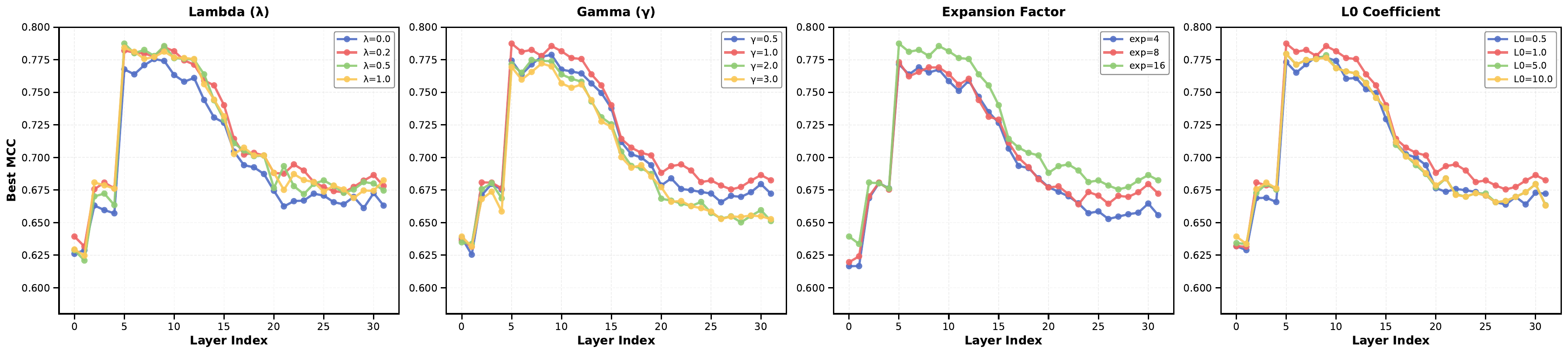}
        \caption{Layer-wise performance trajectory across different hyperparameter settings. The consistent "Inverted-U" pattern validates the structural nature of Signal Submersion.}
        \label{fig:layer_dynamics}
    \end{subfigure}
    \vspace{-3em}
    \caption{Hyperparameter Sensitivity and Layer Dynamics Analysis. All experiments were conducted using the CodeLlama-7B backbone on the BigVul dataset.}
    \label{fig:sensitivity_and_layers}
\end{figure}

\textbf{Hyperparameter Sensitivity \& Structural Invariance.}
We further examine the robustness of the model by adjusting key hyperparameters as shown in Fig. \ref{fig:hyperparam_sensitivity}.
The results indicate that the performance remains highly stable across different settings.
The Sparse Loss ($\lambda$) reaches its optimal performance at 0.5, which confirms that a noise filter is necessary to reduce interference from unrelated features.
Similarly, the Classification Weight ($\gamma$) peaks at 1.0, representing an ideal balance between accurately reconstructing the data and identifying defects.
Additionally, increasing the Expansion Factor up to $16\times$ leads to consistent improvements, validating that projecting features into a larger space is essential for separating the complex logic of vulnerabilities.

Most importantly, Fig. \ref{fig:layer_dynamics} reveals that the signal trajectory consistently follows an inverted-U pattern regardless of the hyperparameter settings.
The detection capability always peaks at the intermediate layers (Layers 4--8) before declining rapidly in the deeper layers.
This empirical evidence supports the observation that vulnerability signals fade as the model processes information.
While the deeper layers of large language models focus on generating text and integrating context, the specific signals required for detecting defects are strongest in the middle layers.
Consequently, SAGE effectively extracts these critical signals before they are overwhelmed by the general objectives of the model.

\vspace{-0.5em}
\begin{boxK}
\small \faIcon{pencil-alt} \textbf{Finding 4 (Synergy and Stability):} Sparsity contributes significantly to achieving high Precision, while task-specific alignment ensures high Recall. SAGE combines these elements to refine the signal more effectively than standard dense methods. Moreover, the consistent inverted-U trajectory across various settings proves that intervening at the intermediate layers is structurally necessary for accurate detection.
\end{boxK}

\subsection{RQ4: Impact of Model Scale}
\label{sec:rq3}

To determine if Signal Submersion stems from limited capacity or architectural bottlenecks, we evaluated CodeLlama (7B-34B) and Qwen2.5-Coder (7B-32B) against SAGE. We designate standard Supervised Fine-Tuning (SFT) models as the Baseline to isolate the impact of mechanistic intervention.

\begin{figure*}[h]
\centering
\includegraphics[width=1\linewidth]{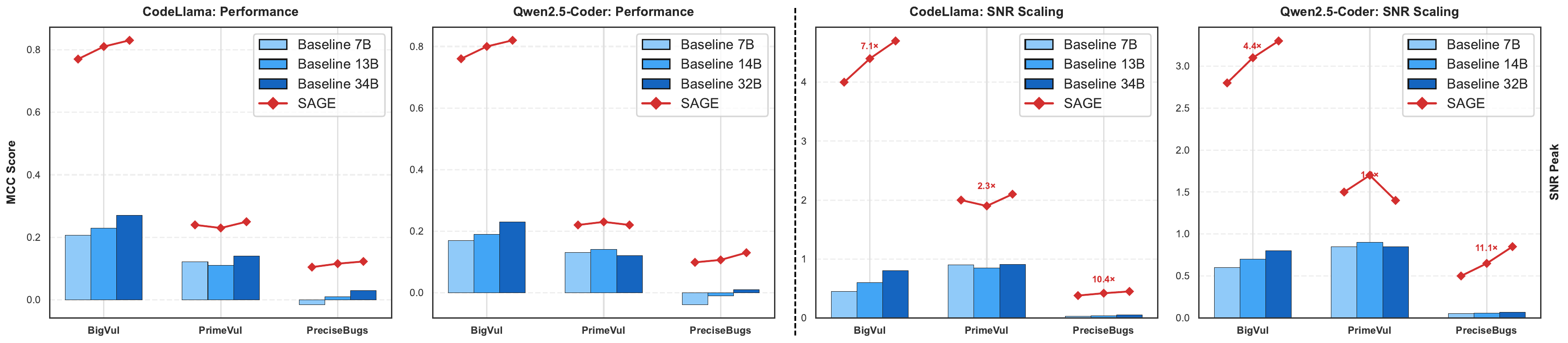}
\caption{Scalability and Signal Quality Analysis. Standard scaling (blue) exhibits diminishing marginal returns in both performance and signal-to-noise ratio, whereas the SAGE algorithm (red) consistently amplifies signals above the noise baseline. Notably, despite a fivefold increase in parameters, the baseline still shows an asymptotic plateau.}
\label{fig:scalability}
\vspace{-0.7em}
\end{figure*}

\textbf{The Saturation of Scaling Laws.}
Our results examine the assumption that simply increasing model size solves the problem of missed detections. As shown in Table~\ref{tab:main_results}, specialized smaller models often outperform massive general-purpose models like GPT-5.2~\cite{gpt52}, Gemini 3~\cite{gemini3}, and Claude 4.5~\cite{claude45}. For instance, SAGE-7B achieves a higher score than Gemini 3 Pro on the BigVul dataset, whereas GPT-5.2 even shows a negative correlation on PreciseBugs. This performance gap suggests that larger models tend to prioritize general fluency and coherence, which causes them to overlook the subtle details associated with vulnerabilities. Fig.~\ref{fig:scalability} further confirms this limitation by revealing a performance plateau as models scale from 7B to 34B parameters. This stagnation indicates that adding more parameters does not necessarily improve the ability of a model to recognize vulnerability features. Consequently, while larger models may store more global patterns, they do not inherently possess the sharp perception required to identify fine-grained security anomalies.

\begin{table*}[h]
\centering
\caption{Efficiency Analysis. SAGE achieves superior efficiency, reducing training time by up to $26\times$ compared to RL methods, despite a higher parameter count due to its frozen backbone architecture.}
\vspace{-1em}
\label{tab:efficiency_horizontal}
\resizebox{1.0\textwidth}{!}{%
\begin{tabular}{l|cccc|cc|ccc}
\toprule
\textbf{Method} & LineVul & VulLLM & MSIVD & ReVD & LineVul & LineVul & \cellcolor{gray!15}\textbf{SAGE} & \cellcolor{gray!15}\textbf{SAGE} & \cellcolor{gray!15}\textbf{SAGE} \\
\textbf{\& Backbone} & (7B) & (7B) & (7B) & (7B) & (13B) & (34B) & \cellcolor{gray!15}\textbf{(7B)} & \cellcolor{gray!15}\textbf{(13B)} & \cellcolor{gray!15}\textbf{(34B)} \\ 
\midrule
\textbf{Params} & $16.8\text{M}$ & $167.7\text{M}$ & $240.0\text{M}$ & $167.7\text{M}$ & $26.2\text{M}$ & $67.1\text{M}$ & \cellcolor{gray!15}$537.0\text{M}$ & \cellcolor{gray!15}$839.0\text{M}$ & \cellcolor{gray!15}$2.15\text{B}$ \\
\textbf{Time (h)} & 13.6 & 34.8 & 40.5 & 69.6 & 25.6 & 62.4 & \cellcolor{gray!15}\textbf{2.6} & \cellcolor{gray!15}\textbf{5.3} & \cellcolor{gray!15}\textbf{15.8} \\ 
\bottomrule
\end{tabular}%
}
\vspace{-1em}
\end{table*}

\textbf{The Cost-Efficiency Paradox \& Frozen Backbone Advantage.}
Standard scaling incurs quadratic computational costs for negligible gains (Table \ref{tab:efficiency_horizontal}). In contrast, \textbf{SAGE-7B outperforms the Baseline-34B} while reducing training latency by \textbf{80.8\%} compared to LineVul-7B and \textbf{26.7$\times$} compared to ReVD. 
Although SAGE possesses more trainable parameters (due to the wide Sparse Autoencoder), it operates on a Frozen Backbone regime. By avoiding expensive backpropagation through the deep transformer layers, SAGE effectively decouples the cost of detection from the cost of reasoning, enabling a standard RTX 3090 to harness the power of 34B-scale representations.

\textbf{Mechanistic Explanation for Frontier Model Failures.}
These findings offer a mechanistic explanation for why massive proprietary models, such as GPT-5.2 and Gemini 3, underperformed in our RQ2 analysis. The observed performance plateau indicates that simply scaling up parameters yields diminishing returns in distinguishing vulnerability signals. This suggests that the issue is not incidental but stems from an architectural limitation: as models grow, their enhanced focus on general functional semantics tends to overshadow the sparse indicators of security vulnerabilities. Consequently, even trillion-parameter models struggle to prioritize these subtle artifacts over global code patterns. Addressing this requires a shift in how models represent data—such as employing sparse projection to isolate relevant features—rather than continuing to rely on increased model size.

\vspace{-0.3em} 
\begin{boxK} 
\small 
\faIcon{pencil-alt} \textbf{Finding 6: The Limitations of Scale.}
We observe a notable gap between model capacity and vulnerability sensitivity. Standard approaches that simply increase model size result in high computational costs with almost no improvement in detection signal quality. SAGE overcomes this limitation by demonstrating that optimizing the model structure is far more effective than relying on parameter scale. Notably, a 7B model equipped with SAGE outperforms a standard 34B baseline while requiring only a fraction of the computational resources.
\end{boxK} 
\vspace{-0.7em}

\section{Implications}

The findings in this work extend beyond the performance improvements of SAGE. They offer important lessons for applying Large Language Models in software engineering. We identified that the main problem is the suppression of weak signals within the model architecture. Therefore, we examine the common assumption that simply increasing the size of a model or using standard post-training methods like reinforcement learning is the best way to improve detection. Instead, we propose a design approach that focuses on optimizing the internal processing mechanisms of the model.

\textbf{Implications for Researchers.}
First, we identify the limits of model scaling and standard training. Many researchers believe that larger models automatically reason better. However, our results show that scaling has a limit. While larger models understand global context, the main features of the code still hide the small signals of a vulnerability. Standard fine-tuning also fails because it only optimizes the final output where the signal is already weak. Researchers should focus on architectures that protect these subtle signals in the intermediate layers.
Second, this work connects model interpretability to performance. Usually, researchers only use interpretability to analyze a model after training. SAGE shows that interpretability tools can actively improve performance during detection. We should stop treating models as a black box systems. Instead, we should use strategies that help the model separate its internal features rather than only checking the final answer.
Third, we show that sparsity is essential for reliability. Standard representations often mix vulnerability logic with irrelevant coding styles. This makes models fail when the style changes. Our results show that simplifying internal representations is necessary for generalization. This allows the model to learn the actual nature of a bug instead of just memorizing the training data.

\textbf{Implications for Practitioners.}
For industry practitioners, our findings highlight a practical opportunity for cost efficiency. We demonstrate that a standard 7B model can outperform larger 34B generalist models by simply attaching a lightweight, external module while keeping the original model's parameters frozen. Since this approach involves very few trainable parameters and does not alter the base model's architecture, it suggests that effective security analysis may not require massive proprietary models or expensive full-model training. Instead, using a small, specialized add-on with a standard model offers a more practical solution for scalable and real-time vulnerability scanning in environments with limited resources.

Furthermore, the study emphasizes the need for stability when facing diverse code. In real-world scenarios, codebases vary greatly and frequently undergo changes or restructuring. Traditional fine-tuning often produces models that perform poorly when facing minor syntactic variations. The ability of SAGE to focus on the consistent core features of a vulnerability provides a guide for building resilient security tools. Practitioners should prioritize models that demonstrate the ability to ignore irrelevant changes. This approach reduces the risk of missing detections when facing new vulnerability patterns or complex code structures.

\section{Conclusion and Future Work}
\label{sec:conclusion}

In this work, we identify Signal Submersion as a significant architectural limitation that limits the ability of Large Language Models to fine-grained security defects, necessitating a shift from passive feature extraction to active signal amplification. By projecting entangled representations onto a high-dimensional sparse manifold, our proposed SAGE framework successfully recovers low-magnitude vulnerability signals, achieving state-of-the-art performance and better robustness against distribution shifts while challenging the universality of parameter scaling for anomaly detection tasks. Looking forward, we plan to extend this method beyond simple detection. Future work will focus on translating these sparse features into natural language to explain the root causes of vulnerabilities. We also aim to use these refined signals to guide models in automatically fixing code. Additionally, we will develop adaptive methods to capture defects at different levels of complexity. Ultimately, SAGE demonstrates that targeted intervention in the internal mechanisms of LLMs offers a more efficient and rigorous path to software security than simply increasing model size.

\section*{Data Availability}
We have open-sourced the code and dataset at: \url{https://github.com/BDS-SDU/SAGE}.

\section*{Acknowledgement}
We thank the anonymous reviewers for their time and helpful comments. This study is supported by the National Natural Science Foundation of China (No. U25A20425, 62232010, 62302266, U23A20302, U24A20244, U24B20149), the Research Project of Quancheng Laboratory, China under Grant No. QCL20250106, the Natural Science Foundation of Shandong Province (Grant No. ZR2024QF093), and the Young Talent of Lifting Engineering for Science and Technology in Shandong, China (Grant No. SDAST2025QTB031)..

\bibliographystyle{ACM-Reference-Format}
\bibliography{reference}

\end{document}